\documentclass[journal]{IEEEtran}
\usepackage{ifpdf}

\usepackage{cite}

\ifCLASSINFOpdf
  \usepackage[pdftex]{graphicx}
\else
  \usepackage[dvips]{graphicx}
\fi
\usepackage{amsmath, amssymb}
\usepackage{algorithmic, algorithm}

\usepackage{array}

\ifCLASSOPTIONcompsoc
 \usepackage[caption=false,font=normalsize,labelfont=sf,textfont=sf]{subfig}
\else
 \usepackage[caption=false,font=footnotesize]{subfig}
\fi

\usepackage{stfloats}
\ifCLASSOPTIONcaptionsoff
 \usepackage[nomarkers]{endfloat}
\let\MYoriglatexcaption\caption
\renewcommand{\caption}[2][\relax]{\MYoriglatexcaption[#2]{#2}}
\fi
\usepackage{url}

\usepackage[colorlinks,linkcolor=red,anchorcolor=green,citecolor=blue]{hyperref}

\usepackage{multirow}
\usepackage{arydshln}
\usepackage[table,xcdraw,dvipsnames]{xcolor}
\usepackage{xspace}
\usepackage{longtable} 
\usepackage{subfig}
\usepackage{subfloat}
\usepackage{booktabs}
\usepackage{threeparttable}
\usepackage{lscape}
\usepackage{verbatim}
\usepackage{marvosym}
\usepackage{ulem}
\usepackage{fancyhdr}
\usepackage{soul}
\usepackage{siunitx}
\usepackage{pifont}

\usepackage{tikz}
\newcommand{\copyrightnotice}{%
  \begin{tikzpicture}[remember picture,overlay]
    \node[anchor=south,yshift=8pt,xshift=0pt] at (current page.south) {%
      \begin{minipage}{0.92\textwidth}
        \footnotesize
        \textcopyright~2026 IEEE. Personal use of this material is permitted. Permission from IEEE must be obtained for all other uses, in any current or future media, including reprinting/republishing this material for advertising or promotional purposes, creating new collective works, for resale or redistribution to servers or lists, or reuse of any copyrighted component of this work in other works.
      \end{minipage}%
    };
  \end{tikzpicture}%
}

\newcommand{\projName}{HiKV\xspace}
\newcommand{\fc}[1]{\textcolor{black}{#1}}

\newcommand{\fcRThree}[1]{\textcolor{black}{#1}} 
\newcommand{\fcRfour}[1]{\textcolor{black}{#1}} 
\newcommand{\fcRfive}[1]{\textcolor{black}{#1}}
\newcommand{\fcRevROne}[1]{\textcolor{black}{#1}}
\newcommand{\fcRevRTwo}[1]{\textcolor{black}{#1}}

\newcommand{\jyinComment}[1]{\textcolor{black}{#1}}

\begin{document}


%

\title{\projName: Hierarchical Importance-Aware KV Cache with Hardware Acceleration for LLM Decoding}

%


\author{Chao~Fang,~\IEEEmembership{Member,~IEEE,}
        Jun~Yin,
        Man~Shi,        
        and~Marian~Verhelst,~\IEEEmembership{Fellow,~IEEE}
\thanks{This work was supported in part by the European Research Council (ERC) under grant agreement No. 101088865, the European Union’s Horizon 2020 program under grant agreement No. 101070374, the Flanders AI Research Program, and long-term structural Methusalem funding by the Flemish Government. \textit{(Corresponding author: Chao Fang.)}}
\thanks{C. Fang, J. Yin, and M. Verhelst are with ESAT-MICAS, KU Leuven, Leuven, Belgium (e-mail: chao.fang@kuleuven.be; jun.yin@kuleuven.be; marian.verhelst@kuleuven.be).}
\thanks{M. Shi was with ESAT-MICAS, KU Leuven, Leuven, Belgium, and is now with UC Berkley, CA, USA (e-mail: manshi@berkeley.edu).}
}
%
%

\markboth{Journal of \LaTeX\ Class Files,~Vol.~14, No.~8, August~2015}%
{Shell \MakeLowercase{\textit{et al.}}: Bare Demo of IEEEtran.cls for IEEE Journals}
%



\maketitle
\copyrightnotice

\begin{abstract}
\fcRevROne{With the rapid adoption of long-context large language models (LLMs), the continuously growing KV cache during decoding has become the critical memory bottleneck.}
\fcRfour{To tackle \fcRevROne{this challenge}, we propose HiKV, a novel algorithm-hardware co-design that exploits KV cache redundancy through hierarchical importance awareness.}
\fcRevROne{Algorithmically, HiKV compresses the KV cache at two granularities: Stage I evicts unimportant tokens within a fixed budget, and Stage II further loads only the significant elements of each retained token, reaching compression ratios unattainable at a single granularity.}
\fcRevROne{Architecturally, we develop a dedicated accelerator centered on a reconfigurable importance sorter that switches between the distinct sorting datapaths each stage requires, unifying the two-stage acceleration in one circuit with minimal overhead.}
\fcRfour{Evaluated on representative LLMs, HiKV achieves up to \fcRevROne{7.95}$\times$ speedup and 90\% energy reduction \fcRfive{in the attention computation }\fcRevROne{over the}
vanilla KV cache baseline within negligible 1\% accuracy loss. Under iso-accuracy constraints, HiKV outperforms state-of-the-art importance-based methods by achieving an additional 1.82$\sim$4.87$\times$ reduction in external memory accesses. These benefits are enabled by specialized hardware components that add only 8\% to the system area.}

\end{abstract}


%
\IEEEpeerreviewmaketitle

\section{Introduction}

\IEEEPARstart{L}{arge} \fcRfour{language models (LLMs)~\cite{jiang2023mistral, qwen2, llama3modelcard, li2023long} have achieved remarkable success across diverse natural language processing tasks~\cite{brown2020language, wang2024mathcoder, bai2024longbench}, from question answering and summarization to code generation and mathematical reasoning.}
\fcRfour{The deployment of LLMs in real-world applications, however, faces significant efficiency challenges, particularly during the decoding phase of autoregressive text generation.}
\fcRfour{In this phase, LLMs generate output tokens sequentially, with each token depending on all previously generated content.}
\fcRfour{This sequential dependency necessitates a key-value (KV) cache mechanism that stores attention states from prior tokens to avoid costly recomputation.}
\fcRfour{While eliminating redundant calculations, the KV cache introduces a critical memory bottleneck that becomes increasingly severe during the decoding phase of LLM inference.}


\begin{figure}[t]
    \centering
    \includegraphics[width=1\columnwidth]{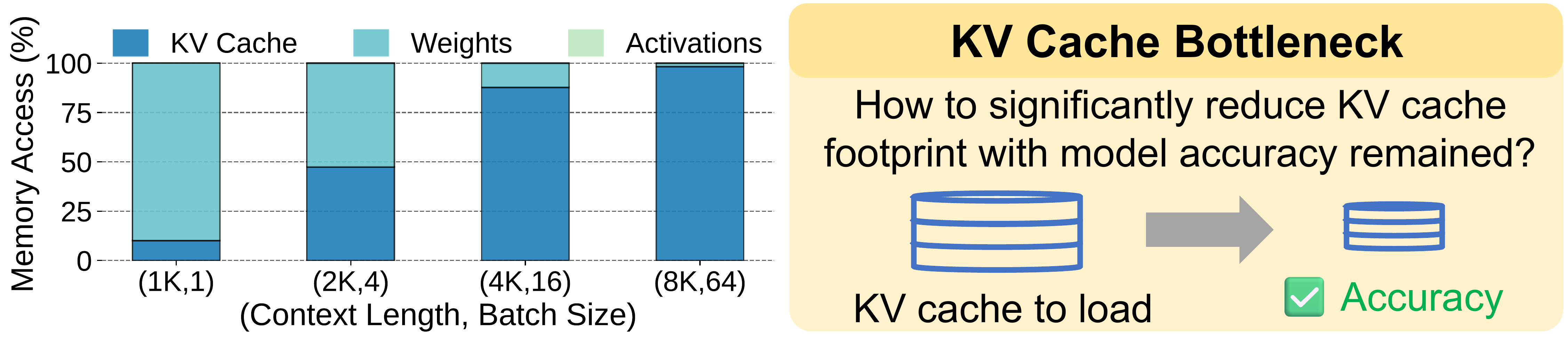}
    \caption{\fcRevROne{The dominated memory access of KV cache becomes the bottleneck of LLM decoding, motivating \projName to shrink its footprint with high accuracy.}}
    \label{fig:motiv}
\end{figure}


\fcRfour{This bottleneck, \fcRevROne{as shown in Fig.~\ref{fig:motiv}}, stems from two compounding factors that simultaneously drive memory footprint growth.}
\fcRfour{First, the cache size grows linearly with sequence length~\cite{chen2025p3, haoyang2025survey, wang2025flashforge}. As LLMs generate longer outputs or process extended contexts for applications like document analysis and multi-turn conversations, each decoding step must access progressively larger cache volumes from external memory.}
\fcRfour{Second, modern serving systems employ batch processing to improve throughput by handling multiple user requests concurrently\cite{yu2022orca, patel2024splitwise, liu2024kivi}. Unlike model weights that can be shared across all requests in a batch, each request must maintain its own independent KV cache across all transformer layers.}
\fcRfour{This dual scaling creates a memory-bound bottleneck where computational units remain underutilized while idling for KV cache transfers, fundamentally limiting inference efficiency.}

\fcRfive{To address this bottleneck, recent research has explored importance-based strategies to compress the KV cache, which can be broadly categorized into two classes. Heuristic methods \cite{xiao2024efficient, xiao2025duoattention, li2024snapkv, chen2025sepllm, han2024lm, beltagy2020longformer} assign token importance statically through predefined rules, such as retaining only the most recent tokens, but often sacrifice accuracy when critical information resides beyond these fixed boundaries.}
\fcRfive{Instead, importance-aware approaches \cite{zhang2023h2o, liu2023scissorhands, li2025kvo, adnan2024keyformer, zhao2024alisa, zhu2025mata, xu2025unicaim, park2024tokenpicker} compute token importance dynamically through attention score accumulation, enabling data-driven eviction that achieves superior accuracy-compression trade-offs. Despite this progress, their notion of importance remains defined exclusively at the token level: once a token is deemed important, its entire key and value vectors are required to load from external memory.}
\fcRfive{As analyzed in Sec.~\ref{sec:algo}, however, KV cache redundancy exists at both token and element levels. Even within retained tokens, only a subset of vector elements significantly influences attention computation.}

\fcRfive{Moreover, dynamically maintaining importance across KV cache tokens at each decoding step introduces substantial sorting overhead, necessitating dedicated hardware support for efficient execution.}
\fcRfive{While hardware-accelerated solutions~\cite{park2024tokenpicker, zhu2025mata, xu2025unicaim, chen2025dias, wang2025veda, zhang2025kv, moradifirouzabadi2025end} have been proposed to mitigate this sorting overhead, the memory bandwidth bottleneck imposed by loading complete key and value vectors for all selected tokens remains unresolved. This points to the need for extending importance-awareness beyond the token level to the elements within each token, motivating a hierarchical approach that systematically eliminates redundancy at both granularities through dedicated hardware support.}
\begin{figure}[t]
    \centering
    \includegraphics[width=\columnwidth]{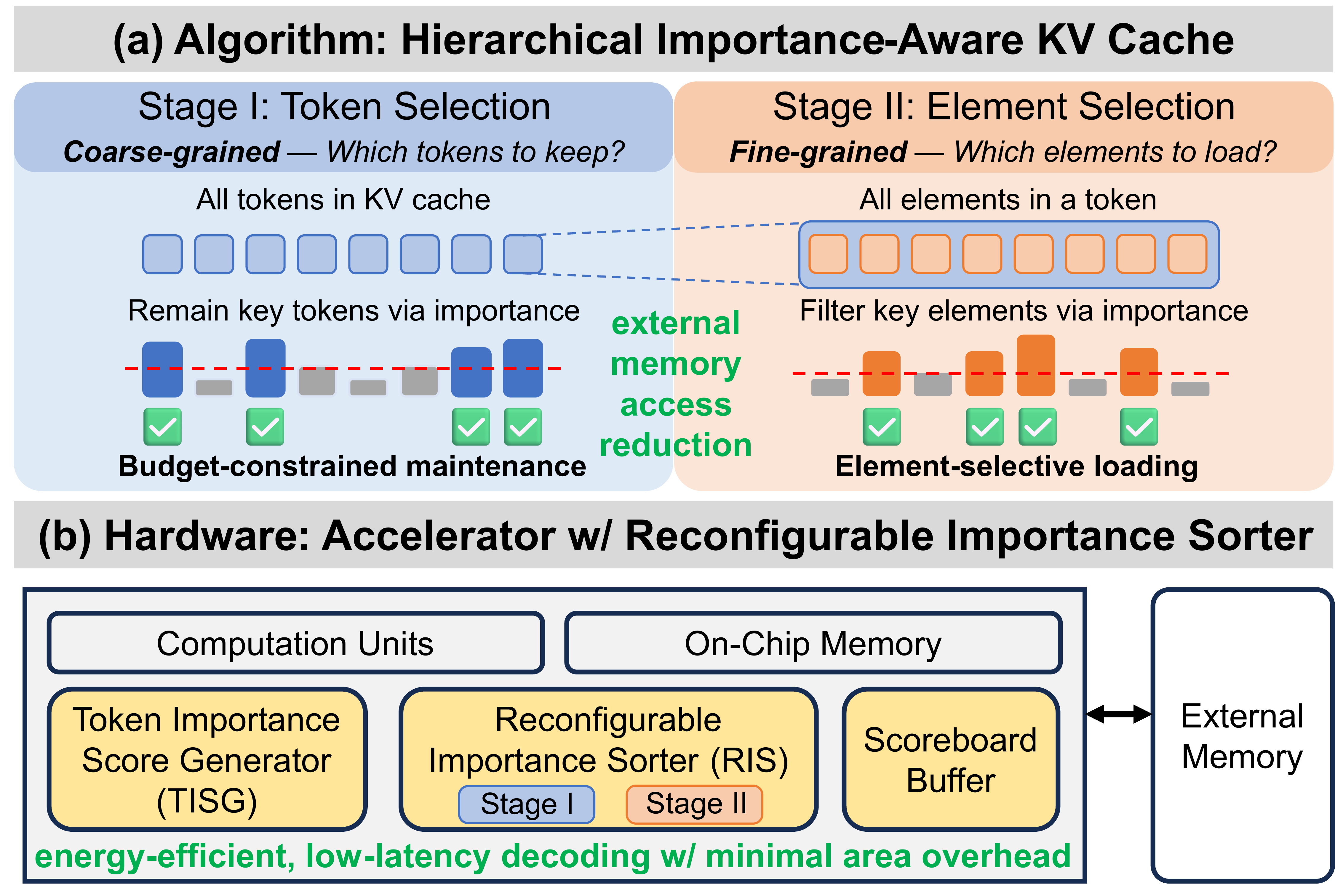}
    \caption{\fcRevROne{Overview of \projName, co-designing (a)~an importance-aware algorithm that compresses the KV cache at both the token and element levels and (b)~an accelerator featuring a reconfigurable importance sorter (RIS), achieving up to 7.95$\times$ speedup and 90\% energy reduction with only 8\% area overhead.}}
    \label{fig:overview}
\end{figure}

\fcRfive{In this paper, we propose \projName, as presented in \fcRevROne{Fig.~\ref{fig:overview}}, an algorithm-hardware co-design that addresses the KV cache bottleneck through hierarchical importance-aware optimization. Algorithmically, \projName operates through two complementary stages: Stage I identifies and retains only important tokens within a fixed cache budget, while Stage II selectively loads only significant vector elements from retained tokens, together achieving compression ratios unattainable by any single-granularity approach. Architecturally, \projName introduces a custom accelerator featuring a reconfigurable sorter that efficiently supports both stages: heap-based importance maintenance in Stage I and parallel chunk-based sorting in Stage II, delivering substantial performance and energy improvements with minimal area overhead.}

\fcRfour{In summary, the main contributions of this paper are:}
\begin{itemize}
    \item \fcRfour{A hierarchical importance-aware approach that exploits KV cache redundancy at both token-level and element-level granularities, achieving superior compression efficiency over token-level state-of-the-art (SotA) methods within 1\% accuracy loss across representative LLMs on long-context benchmarks. (Sec.~\ref{sec:algo})}
    \item \fcRfour{A customized hardware accelerator featuring a reconfigurable sorter that efficiently supports both stages through heap-based importance maintenance for token-level optimization and chunk-based parallel sorting for element-level selection, achieving low overhead with only 8\% system-level area increase. (Sec.~\ref{sec:arch})}
    \item \fcRfour{Comprehensive evaluation \jyinComment{results} demonstrating up to \fcRevROne{7.95}$\times$ speedup and 90\% energy reduction in \fcRfive{the attention computation during LLM decoding} in contrast with the vanilla baseline, outperforming SotA KV cache optimizations under iso-accuracy constraints. (Sec.~\ref{sec:eval})}
\end{itemize}

\fcRfour{The rest of the paper is organized as follows. Sec.~\ref{sec:bkg} presents the background and motivation for KV cache optimization. Sec.~\ref{sec:algo} details the proposed two-stage importance-aware KV cache optimization approach. Sec.~\ref{sec:arch} elaborates on the \projName hardware architecture that enables efficient hierarchical importance maintenance. Sec.~\ref{sec:eval} 
\jyinComment{evaluates our approach in terms of algorithm-, hardware-, and system-level performance.}
}
\fcRfour{Finally, Sec.~\ref{sec:concls} concludes the paper.}

\section{Background and Motivation} 
\label{sec:bkg}

\begin{figure}[t]
    \centering
    \includegraphics[width=1\columnwidth]{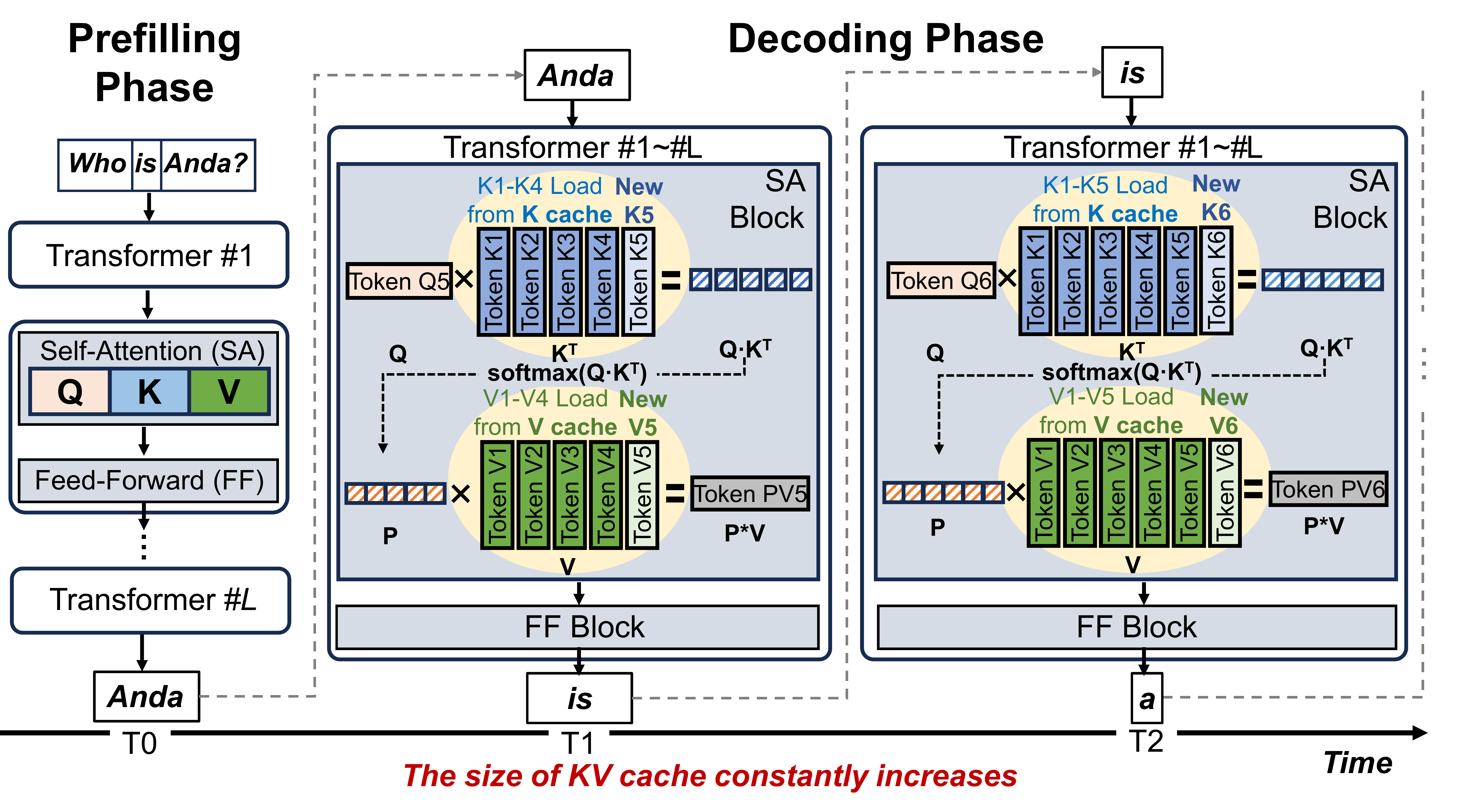}
    \caption{\fcRevROne{During LLM autoregressive decoding, the KV cache grows linearly with each generation step, continuously amplifying external memory access volume and imposing an escalating memory bandwidth bottleneck.}}
    \label{fig:llm}
\end{figure}

\subsection{LLM Generation Fundamentals}
As shown in Fig.~\ref{fig:llm}, LLMs execute generation tasks through a prefill 
phase followed by a decoding phase~\cite{huang2025edgellm, zhong2024distserve, zeng2024flightllm, wang2025ofq}.
During the prefill phase at timestamp T0, the model processes all input token vectors of the prompt simultaneously, computing query (Q), key (K) and value (V) matrices via self-attention mechanisms.
These K and V matrices are stored in a structure, namely the KV cache~\cite{yuan2024kvcache, ma2025apt, zhou2024survey}, to avoid costly recomputation in subsequent processing.
In the decoding phase, new tokens are generated step by step at timestamps T1, T2, ..., until the maximum sequence length is reached or an end-of-sequence token appears~\cite{guo2024stop}.
\fc{At each timestamp, the model performs two critical attention computations \fcRfive{in general matrix-vector (GeMV) operations}: $QK^T$ determines token relationships and $PV$ generates new context representations, where $P=\text{softmax}(QK^T)$.}
As text decoding proceeds, each new key-value token is continuously appended to the KV cache, causing linear expansion in memory usage.
Unlike the prefill phase, where computation can be parallelized, the sequential nature of decoding, combined with the growing KV cache access demands, introduces severe computational challenges in LLM decoding.

\subsection{KV Cache Bottleneck}

In LLM decoding, memory access patterns significantly affect both performance and energy efficiency, particularly in the memory-bound decoding phase~\cite{zhang2024llmcompass, geens2024energy, yuan2024llm}.
LLM services typically leverage batch processing~\cite{yu2022orca, patel2024splitwise, liu2024kivi} to handle multiple user requests concurrently and improve computational efficiency.
Yet, this batching strategy's impact on memory access patterns needs careful examination.

\begin{figure}[t]
\centering
\includegraphics[width=\columnwidth]{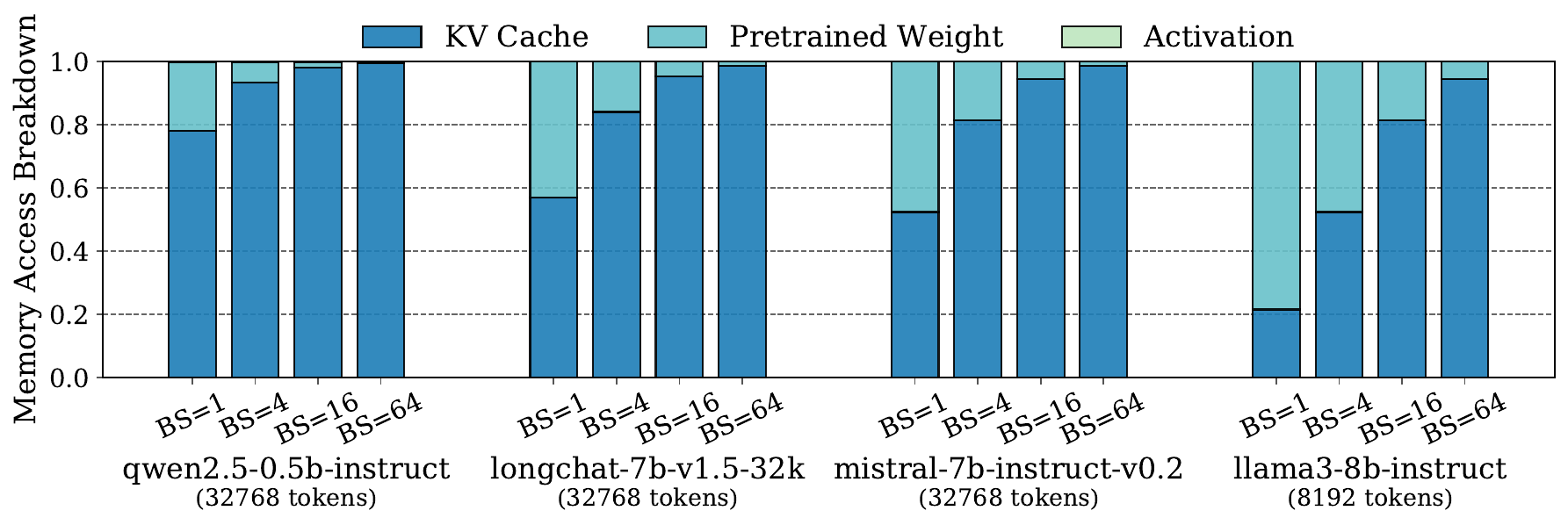}
\caption{\fcRfive{Model memory transfer breakdown with different batch sizes when decoding tokens at long-context scenarios. KV cache increasingly dominates as batch size grows, while activation remains negligible.}}
\label{fig:mem_comp}
\end{figure}

Using the LLM-Viewer~\cite{yuan2024llm} tool, we visualize the memory transfer breakdown of four popular LLMs~\cite{qwen2, li2023long, jiang2023mistral, llama3modelcard} when generating sequences 
\fcRfive{in long-context scenarios} under various batch size requests.
As shown in Fig.~\ref{fig:mem_comp}, the KV  cache dominates memory accesses compared to other components, i.e., pre-trained weights and activations, especially as batch size increases.
For instance, in llama3-8b-instruct model~\cite{llama3modelcard}, KV cache consumes around 20\% of memory access within a single batch but escalates sharply to over 90\% at a batch size of 64.
This dramatic increase occurs because each request in a batch requires its own KV cache in every attention layer while model weights can be shared across requests.
These dominant off-chip memory accesses not only incur high latency but also consume significant energy, making KV cache optimization crucial for efficient LLM decoding.

\subsection{KV Cache Compression Techniques}

\fcRfour{Given the memory bottleneck discussed above, recent research has explored various importance-based strategies to reduce KV cache footprint, which can be broadly divided into two types. Heuristic methods~\cite{xiao2024efficient, xiao2025duoattention, li2024snapkv, chen2025sepllm, han2024lm, beltagy2020longformer} employ predefined eviction policies without dynamic importance computation. StreamingLLM~\cite{xiao2024efficient} retains only recent tokens and a fixed number of initial attention sink tokens, avoiding expensive scoring operations. DuoAttention~\cite{xiao2025duoattention} separates attention heads into retrieval and streaming types. SepLLM~\cite{chen2025sepllm} segments sequences into compressed representations, and Longformer~\cite{beltagy2020longformer} introduces local-global attention patterns. However, these rigid windowing or pattern-based strategies often sacrifice accuracy in long-context scenarios where important information may reside beyond predefined boundaries.}

\fcRfour{To solve the issues, importance-aware methods~\cite{zhang2023h2o, liu2023scissorhands, li2025kvo, adnan2024keyformer, zhao2024alisa, zhu2025mata, xu2025unicaim, park2024tokenpicker, tian2025skipkv} dynamically track token significance through attention score accumulation. H2O~\cite{zhang2023h2o} maintains a global importance score for each token by accumulating attention weights across all decoding steps, evicting tokens with the lowest scores when cache capacity is exceeded. Scissorhands~\cite{liu2023scissorhands} exploits the persistence of importance across layers, KVO-LLM~\cite{li2025kvo} optimizes importance tracking for batched inference, Keyformer~\cite{adnan2024keyformer} selects key tokens for efficient generation, and ALISA~\cite{zhao2024alisa} accelerates inference through sparsity-aware caching. While these approaches demonstrate superior accuracy-compression trade-offs compared to heuristic methods, they introduce substantial sorting overhead: maintaining importance scores across potentially thousands of tokens at each decoding step becomes a performance bottleneck.}

\subsection{Hardware-Accelerated KV Cache Processing}

\fcRfour{Recognizing the sorting overhead introduced by importance-aware methods, prior arts~\cite{park2024tokenpicker, zhu2025mata, xu2025unicaim, chen2025dias, wang2025veda, zhang2025kv, chen2025titanus, moradifirouzabadi2025end} have explored dedicated hardware solutions to accelerate KV cache management. Token-Picker~\cite{park2024tokenpicker} proposes a dedicated hardware accelerator that prunes redundant KV token accesses through probability estimation, incorporating specialized sorting units to reduce token selection latency. However, Token-Picker still operates exclusively at the token-level granularity: once a token is selected, its entire vector representation must be loaded from external memory. Consequently, despite eliminating over half of the V cache accesses, it requires loading a substantial portion of the K cache, limiting overall memory bandwidth reduction.}
\fcRfour{MATA~\cite{zhu2025mata} designs memory-efficient attention through look-back KV cache pruning. UniCAIM~\cite{xu2025unicaim} proposes a unified architecture with static-dynamic cache pruning. DIAS~\cite{chen2025dias} exploits distance-based attention sparsity with tree-like processing-in-memory. VEDA~\cite{wang2025veda} enables efficient generation through voting-based eviction and dataflow-flexible acceleration. Moreover, Titanus~\cite{chen2025titanus} enables KV cache pruning and quantization on-the-fly.}


\fcRfive{While these works demonstrate the value of specialized hardware for KV cache optimization, none systematically combine hierarchical importance-aware optimization across both token and element granularities with unified hardware support. This motivates our hierarchical approach that eliminates redundancy at multiple granularities through a two-stage importance-aware mechanism with dedicated hardware acceleration.}
\section{Hierarchical Importance-Aware KV Cache} \label{sec:algo}

\fcRfour{This section elaborates on the proposed HiKV technique, a hierarchical importance-aware KV cache. We first analyze the distribution patterns of attention scores to reveal optimization opportunities, then introduce the algorithm designs at both coarse-grained and fine-grained levels, and finally demonstrate how they integrate into the LLM decoding pipeline.}


\begin{figure}[t]
\centering
\includegraphics[width=1.00\columnwidth]{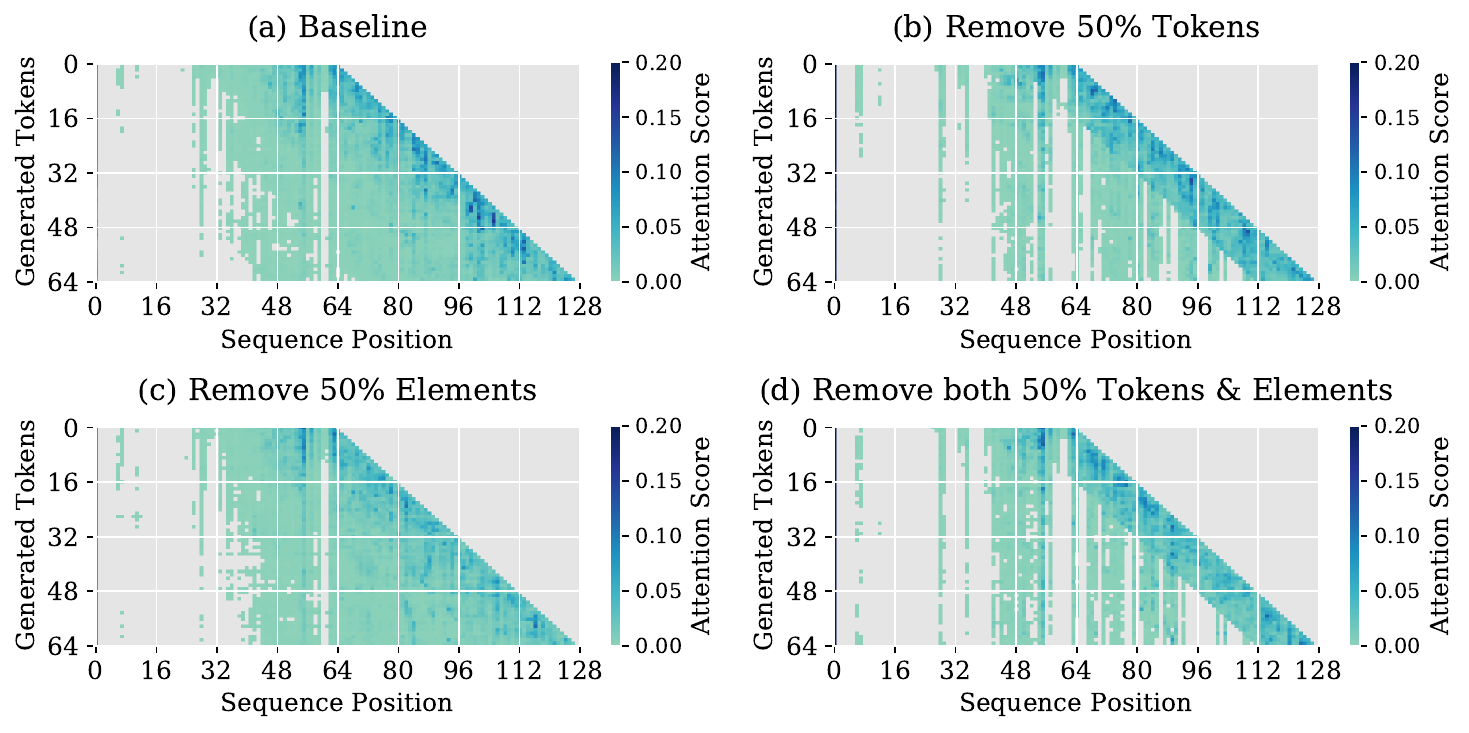}
\caption{Visualization of the average attention scores of qwen2.5-0.5b~\cite{qwen2} for step-by-step generated tokens 
with zero-attention regions in gray: 
(a) baseline; (b) removing 50\% unimportant tokens; (c) removing 50\% unimportant elements per token; (d) removing both 50\% unimportant tokens and elements.}
\label{fig:attn_heatmep}
\end{figure}

\subsection{Importance Analysis of KV Cache}
Recent studies~\cite{xiao2024efficient, zhang2023h2o, lee2024infinigen} have demonstrated that KV cache tokens exhibit distinct importance to token generation accuracy. Motivated by this observation, we conduct a comprehensive analysis to further explore potential optimization opportunities. 
Attention scores $P$, where $P = \text{softmax}(QK^T)$, naturally reflect the contribution of each token during \fcRfour{the LLM decoding process~\cite{ribar2024sparq}}.
Using this metric, we first analyze the attention score distribution patterns of qwen2.5-0.5b~\cite{qwen2} model in the NarrativeQA task~\cite{bai2024longbench} with a simple case. It is with 8 input sequences, each consisting of 64 prefilled tokens and 64 generated tokens.
As shown in Fig.~\ref{fig:attn_heatmep}, our analysis uncovers optimization opportunities 
not only \fcRfour{between tokens but also within individual token elements.}

Fig.~\ref{fig:attn_heatmep}(a) shows attention scores are non-uniform, with certain tokens receiving higher attention weights in dark blue regions while others have minimal impact.
Furthermore, as depicted in Fig.~\ref{fig:attn_heatmep}(b), when removing 50\% of input K cache tokens with the lowest attention scores, the primary attention patterns in the dark blue region remain preserved, \fcRfour{demonstrating that a substantial portion of tokens contribute minimally to attention computation and can be safely excluded from the cache.}
\fcRevROne{While prior importance-aware methods~\cite{xiao2024efficient, zhang2023h2o, zhu2025mata} exploit only this between-token redundancy and stop at the token level, we further observe substantial redundancy within the elements of each token.}
As shown in Fig.~\ref{fig:attn_heatmep}(c), preserving only the top 50\% elements of the input Q token, measured by their magnitude, maintains the overall attention distribution pattern, \fcRfour{indicating that many elements within each token have negligible impact on attention scores}.
Most importantly, these two levels of optimization can be orthogonally combined for higher compression ratios.
Fig.~\ref{fig:attn_heatmep}(d) demonstrates that applying both 50\% compression at \fcRfour{token and element levels},
which achieves 4$\times$ total compression ratio, still preserves key attention patterns.
These observations reveal that we can only load and compute important portions of KV cached tokens
to avoid redundant memory accesses and computations, motivating our \fcRfour{proposed hierarchical two-stage approach to KV cache management}.

\subsection{Stage I: Coarse-Grained Token--Level Management} \label{subsec:algo_stagei}
\fcRfour{To exploit redundancy between tokens, Stage I implements a token-level management strategy for KV cache. It retains only important tokens within a fixed cache budget, thereby reducing the number of tokens accessed from external memory during decoding.}

Given a target compression ratio $r_{I}$ of the KV cache and prefilling sequence length $L_{p}$, we establish a fixed KV cache budget as $B_{I} = L_{p} / r_{I}$.
The attention score vector P quantifies the importance of prior tokens in generating the current token.
\fcRevROne{The key challenge is to track token importance accurately while keeping the overhead low.}
\fcRevROne{Existing token-level methods~\cite{xiao2024efficient, zhang2023h2o, zhu2025mata} achieve only one. Static rules~\cite{xiao2024efficient} are cheap but drop later-important tokens, while dynamic methods like H2O~\cite{zhang2023h2o} stay accurate but re-sort all tokens at every step.}

\begin{figure}[t]
\centering
\includegraphics[width=\columnwidth]{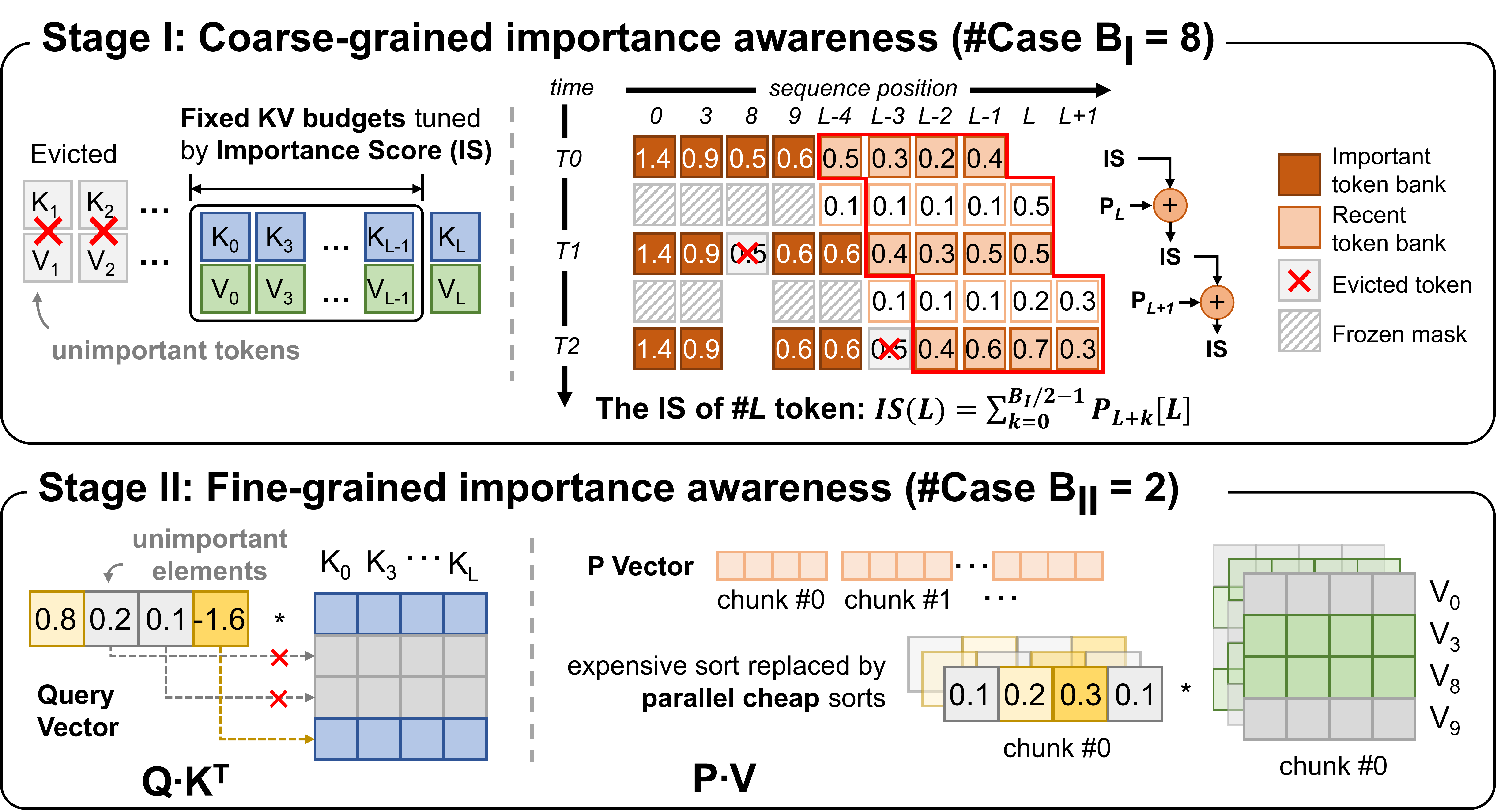}
\caption{\fcRevROne{\projName applies orthogonal compression at two independent granularities at each decoding step. Stage I maintains a dual-bank structure to evict unimportant tokens, reducing the number of tokens loaded. Stage II selects significant elements via Q and P vector sorting, reducing the volume loaded per token. The two stages act on different dimensions of the KV cache and are jointly formalized in Algorithm~\ref{algo:hikv}.}}
\label{fig:algo_flow}
\end{figure}

\fcRfour{Instead, we propose a localized accumulation strategy that exploits token IS more efficiently. The budget $B_I$ is partitioned equally into a recent token bank $\mathcal{R}$ and an important token bank $\mathcal{I}$. Notably, $\mathcal{R}$ stores the latest consecutive tokens of the generation sequence, while $\mathcal{I}$ stores historically significant tokens excluding those in $\mathcal{R}$.}
Rather than global updating IS in H2O~\cite{zhang2023h2o}, \fcRfour{we only accumulate P vectors within $\mathcal{R}$.}
\fcRevROne{As shown in the upper part of Fig.~\ref{fig:algo_flow}, for a token at position $\ell$ in the recent bank $\mathcal{R}$, we compute its IS by accumulating attention weights across the $B_I/2$ decoding steps from its generation step onward: $\text{IS}(\ell) = \sum_{k=0}^{B_I/2 - 1} P_{\ell+k}[\ell]$, where $P_{\ell+k}[\ell]$ is the attention weight received by token $\ell$ at step $\ell+k$.}

\fcRfour{The upper part of Fig.~\ref{fig:algo_flow} further presents the accumulation process across three consecutive timestamps T0, T1, and T2, showing how IS evolves for tokens in the recent bank.}
When a new token is generated and fed into the recent token bank, the oldest token in this bank is ejected and evaluated. If its IS exceeds the minimum IS in the important bank, the token is promoted to the important token bank, evicting the token with the lowest IS there. Otherwise, it is also evicted from the important token bank.
Notably, all the IS in the important token bank are frozen to enable efficient eviction maintenance with the use of a min-heap data structure~\cite{williamsalgorithm}, greatly reducing token eviction complexity from $O(B_{I})$ to $O(log B_{I}$).
\fcRevROne{This freezing keeps the maintenance efficient with minor effect on adaptivity. Within a single-turn generation, the persistence of importance~\cite{liu2023scissorhands} makes a token losing relevance after being important uncommon, while across multi-turn conversations HiKV recomputes the IS of all prior tokens at each turn boundary to demote stale tokens and re-admit revived ones.}

\subsection{Stage II: Fine-Grained Element-Level Management} \label{subsec:algo_stageii}
\fcRfour{Building on the observation that many elements within tokens have negligible impact on attention scores, Stage II further reduces memory transfer volume through element-level selection.}
\fcRevROne{Unlike prior token-level methods~\cite{xiao2024efficient, zhang2023h2o, zhu2025mata} that treat a token as an indivisible unit and must fetch its complete K/V vector once retained, Stage II selects only the significant elements within each retained token.}
\fcRfour{The key insight stems from the asymmetric scale of memory access in attention computation: during decoding, Q and P are single vectors whose memory footprint is minimal, while K and V remain as matrices stored in external memory. Even with batch processing, K and V caches are not shared across requests, making matrix access the dominant memory bottleneck. Hence, we leverage the element magnitudes of Q and P to identify which corresponding positions in K and V matrices are significant, enabling selective loading that substantially reduces memory transfer volume.}

\fcRfour{We define a selected element budget $B_{II} = d_h/r_{II}$, where $d_h$ denotes the attention head dimension and $r_{II}$ is the target element-level compression ratio.}
This determines the number of elements to retain per $Q$ token for $QK^T$ computation. 

\begin{figure}[t]
    \centering
    \includegraphics[width=\columnwidth]{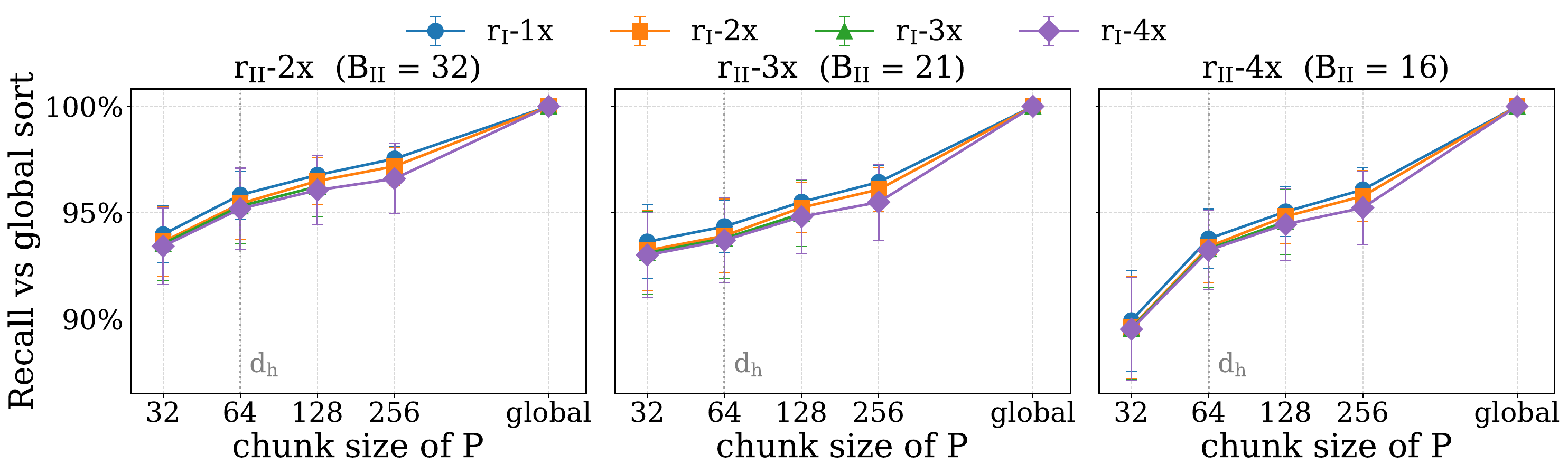}
    \caption{\fcRevROne{Chunk-based sorting recall on Qwen2.5-0.5B-Instruct, evaluated on 8 samples from the 2wikimqa task in LongBench. Each point is the mean over all query heads, layers, and samples. Error bars show the standard deviation.}}
    \label{fig:r46}
    \vspace{-0.3cm}
\end{figure}

Yet, it is important to notice that the $Q$ token and $P$ vector require different sorting strategies to determine their most significant elements, stemming from their distinct characteristics as shown in the lower part of Fig.~\ref{fig:algo_flow}.
\fcRevROne{Formally, let $\mathrm{Top}_{B_{II}}(\cdot)$ return the indices of the $B_{II}$ largest-magnitude entries of a vector.}
Since each $Q$
token dimension equals the model's head dimension $d_h$, which is typically limited to 64 or 128, global sorting is feasible in an efficient way.
\fcRevROne{We thus select a single index set $\mathcal{E}_Q=\mathrm{Top}_{B_{II}}(Q)$ over the whole $Q$ token, which guides the loading of the corresponding $K$ cache for $QK^T$.}
$P$ vectors, however, pose a greater challenge as their length can reach thousands of elements even after Stage I optimization, making global sorting computationally prohibitive.
To address this challenge, as shown in the bottom of Fig.~\ref{fig:algo_flow}, we introduce a chunk-based sorting strategy that turns the expensive global sorting into multiple independent local sortings.
\fcRevROne{Specifically, let $N$ be the length of the $P$ vector after Stage~I. We partition $P$ into $C=\lceil N/d_h\rceil$ chunks of size $d_h$ matching the $Q$ token dimension, and select within each chunk independently, $\mathcal{E}_P^{(c)}=\mathrm{Top}_{B_{II}}\big(P^{(c)}\big)$ for $c=1,\dots,C$, where $P^{(c)}$ is the $c$-th chunk. The selected indices $\{\mathcal{E}_P^{(c)}\}_{c=1}^{C}$ then guide the loading of the corresponding $V$ cache for $PV$.}
\fcRevRTwo{On the V side, since each position in P corresponds to a specific token, the selected elements are chunk-local token rows rather than sub-token dimensions.}
\fcRevROne{This chunk-based design reduces both memory access and computation. Since only $B_{II}=d_h/r_{II}$ of the $d_h$ elements \fcRevRTwo{(i.e., token rows)} per token are loaded and computed, both the $V$ cache memory access and the $PV$ computation drop by a factor of $r_{II}$, and the selection cost of the $P$ vector is lowered from $O(N\log N)$ for a global sort to $O(N\log d_h)$ across $C$ parallelizable chunks.}

\fcRevROne{To verify that this approximation is near-lossless, we compare the elements selected by chunk-based sorting against those selected by a global sort over the full $P$ vector on 8 samples from Qwen2.5-0.5B-Instruct, varying $r_I$ and $r_{II}$ from $1{\times}$ to $4{\times}$. As shown in Fig.~\ref{fig:r46}, for example, at $r_I{=}4{\times}, r_{II}{=}2{\times}$ with chunk size equal to $d_h$, the chunk-based selection achieves over 95\% recall, i.e., it retains over 95\% of the elements that a global sort would select. The overlap remains stable across all tested $r_I$ values, indicating that the chunking error is independent of the Stage~I compression ratio.}

\begin{algorithm}[t]
\small
\caption{\small HiKV Attention Computation at Decoding Step $t$}
\label{algo:hikv}
\begin{algorithmic}[1]
\REQUIRE Query $Q_t$, KV cache ($\mathcal{R}$: Recent bank, $\mathcal{I}$: Important bank)
\ENSURE Attention output $\text{O}_t$, KV cache with updated $\mathcal{R}$ and $\mathcal{I}$
\STATE \textcolor{blue}{// $QK^T$ Computation with hierarchical selection}
\STATE \fcRfive{$\text{tokens} \leftarrow \mathcal{R} \cup \mathcal{I}$ \textcolor{YellowOrange}{$\triangleright$ Stage I}}
\STATE \fcRfive{$\text{elements}_Q \leftarrow$ Select top-$B_{II}$ indices of $Q_t$ \textcolor{Salmon}{$\triangleright$ Stage II}}
\STATE \fcRfive{Load $K[\text{tokens}, \text{elements}_Q]$ from external memory\fcRevRTwo{, where each element in $\text{elements}_Q$ is a head dimension}}
\STATE $P_t \leftarrow \text{Softmax}(Q_t \cdot K^T)$
\STATE \textcolor{blue}{// Maintenance for $\mathcal{R}$ and $\mathcal{I}$}
\STATE \fcRevROne{$\text{IS}[\ell] \leftarrow \text{IS}[\ell] + P_t[\ell],\ \forall \ell \in \mathcal{R}$ \textcolor{YellowOrange}{$\triangleright$ Stage I}}
\STATE \fcRevROne{$\text{evi} \leftarrow$ oldest token in $\mathcal{R}$ \textcolor{YellowOrange}{$\triangleright$ Stage I}}
\STATE \fcRevROne{Promote $\text{evi}$ to $\mathcal{I}$ if $\text{IS}[\text{evi}] > \min\{\text{IS}[j] : j \in \mathcal{I}\}$ \textcolor{YellowOrange}{$\triangleright$ Stage I}}
\STATE \textcolor{blue}{// $PV$ Computation with hierarchical selection}
\STATE Partition $P_t$ into $C$ chunks
\FORALL{chunk $c \in \{1\ldots C\}$}
    \STATE \fcRfive{$\text{elements}_c \leftarrow$ Select top-$B_{II}$ indices from chunk $c$ \textcolor{Salmon}{$\triangleright$ Stage II}}
\ENDFOR
\STATE \fcRfive{Load $V[\text{tokens}, \{\text{elements}_c\}_{c=1}^C]$ from external memory\fcRevRTwo{, where each element in $\text{elements}_c$ is a token row.}}
\STATE $\text{O}_t \leftarrow P_t \cdot V$
\STATE Append $(K_t, V_t)$ of newly generated token to $\mathcal{R}$
\RETURN $\text{O}_t$, Updated $\mathcal{R}$ and $\mathcal{I}$
\end{algorithmic}
\end{algorithm}

\subsection{Hierarchical Decoding Integration}
\label{subsec:integration}

\fcRfour{Having established the individual mechanisms of token-level and element-level optimization, we elaborate how these two stages collaborate throughout the LLM decoding pipeline to systematically reduce KV cache memory access overhead.}

\fcRfour{Before the decoding phase starts, HiKV leverages the attention scores generated during the prefilling phase to initialize the dual-bank structure.}
\fcRfour{The last $B_I/2$ consecutive tokens from the prefill sequence are directly placed into the recent bank $\mathcal{R}$, while from the remaining $L_{p} - B_I/2$ tokens, the top $B_I/2$ tokens, with the highest 
\fcRfive{ISs following the IS definition in Sec.~\ref{subsec:algo_stagei},}
populate the important bank $\mathcal{I}$. \fcRevROne{Since prefill processes the prompt in parallel, each prefill token's IS is computed in one pass by summing its attention column over the $B_I/2$-step window.} This one-time initialization incurs minimal overhead, as dual-bank construction requires only an additional $O(L \log L)$ sorting operation, negligible compared to prefill's inherent $O(L^2)$ attention computation~\cite{zhong2024distserve}, and prefill itself constitutes a small fraction of total inference time in long-context scenarios.}

\begin{figure}[t]
\centering
\includegraphics[width=0.92\columnwidth]{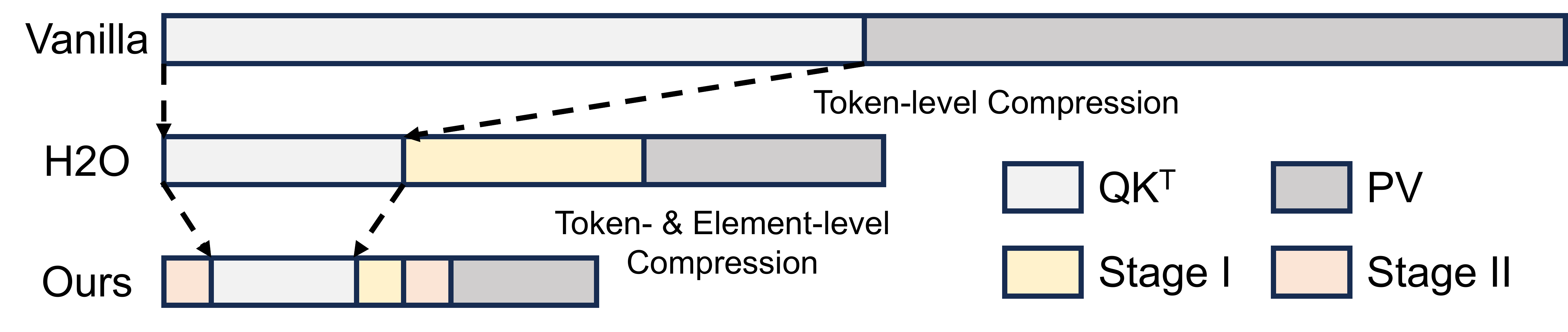}
\caption{\fcRevROne{Operation flow of one decoding step for vanilla, H2O~\cite{zhang2023h2o}, and HiKV.}}
\label{fig:timeline}
\vspace{-0.3cm}
\end{figure}

\fcRfour{During the decoding phase, Algorithm~\ref{algo:hikv} presents the hierarchical optimization workflow for attention computation executed at each token generation step.}
\fcRevROne{The process begins with $QK^T$ computation. Stage~I determines the token set $\mathcal{R}\cup\mathcal{I}$ (Line~2) and Stage~II identifies the top-$B_{II}$ element indices of $Q_t$ by magnitude, producing $\text{elements}_Q$ (Line~3). The system selectively loads $K[\text{tokens}, \text{elements}_Q]$ from external memory (Line~4) and computes $P_t$ via softmax (Line~5). The dual-bank maintenance then updates ISs in $\mathcal{R}$ using $P_t$ (Line~7) and evicts or promotes the oldest token against $\min(\mathcal{I})$ via heap comparison (Line~8--9). For $PV$ computation, Stage~II partitions $P_t$ into $C$ chunks and selects the top-$B_{II}$ elements \fcRevRTwo{(i.e., token rows)} from each (Line~11--13), guiding selective V-cache loading (Line~15). The output $O_t$ is computed (Line~16) and $(K_t, V_t)$ is appended to $\mathcal{R}$ (Line~17).}
\fcRevROne{Fig.~\ref{fig:timeline} complements Algorithm~\ref{algo:hikv} with an operation-flow view contrasting HiKV against vanilla and H2O~\cite{zhang2023h2o} decoding, where bar width reflects the relative number of operations and memory accesses. The $QK^T$ and $PV$ bars shrink progressively from vanilla to H2O to HiKV, reflecting token-level compression in H2O and the additional element-level compression in HiKV. Two Stage~II sort bars appear in the HiKV row for K-side and V-side element selection, and the Stage~I sort bar captures the token eviction overhead, which the min-heap structure reduces to $O(\log B_I)$, making it noticeably narrower than the $O(B_I)$ Stage~I bar in H2O. Stage~II therefore appears twice as the two halves of a single element-level selection, with the K-side half running before the Stage~I eviction. This ordering follows from data dependency rather than the stage indices, as the K-side selection needs only $Q_t$ available at the step start, whereas the Stage~I eviction and the V-side selection both consume $P_t$ produced by $QK^T$.}
\fcRfour{To fully realize these algorithmic benefits, Sec.~\ref{sec:arch} elaborates a custom accelerator design featuring specialized hardware support for various sorting operations in Stage~I and Stage~II.}
\section{\projName Accelerator Architecture} \label{sec:arch}
\fcRfour{This section presents the hardware architecture of the \projName accelerator. The overall architecture centers around the reconfigurable importance sorter (RIS), which flexibly switches between heap-based operations for Stage~I and chunk-based parallel sorting for Stage~II, jointly supporting the two-stage execution flow of hierarchical KV cache optimization. We detail the system architecture overview, the RIS circuit design, and the RIS operations in Stage~I and Stage~II\fcRevROne{, and finally the computing datapath and external memory layout that complete the execution pipeline}.}
 
\begin{figure}[t]
\centering
\includegraphics[width=0.9\columnwidth]{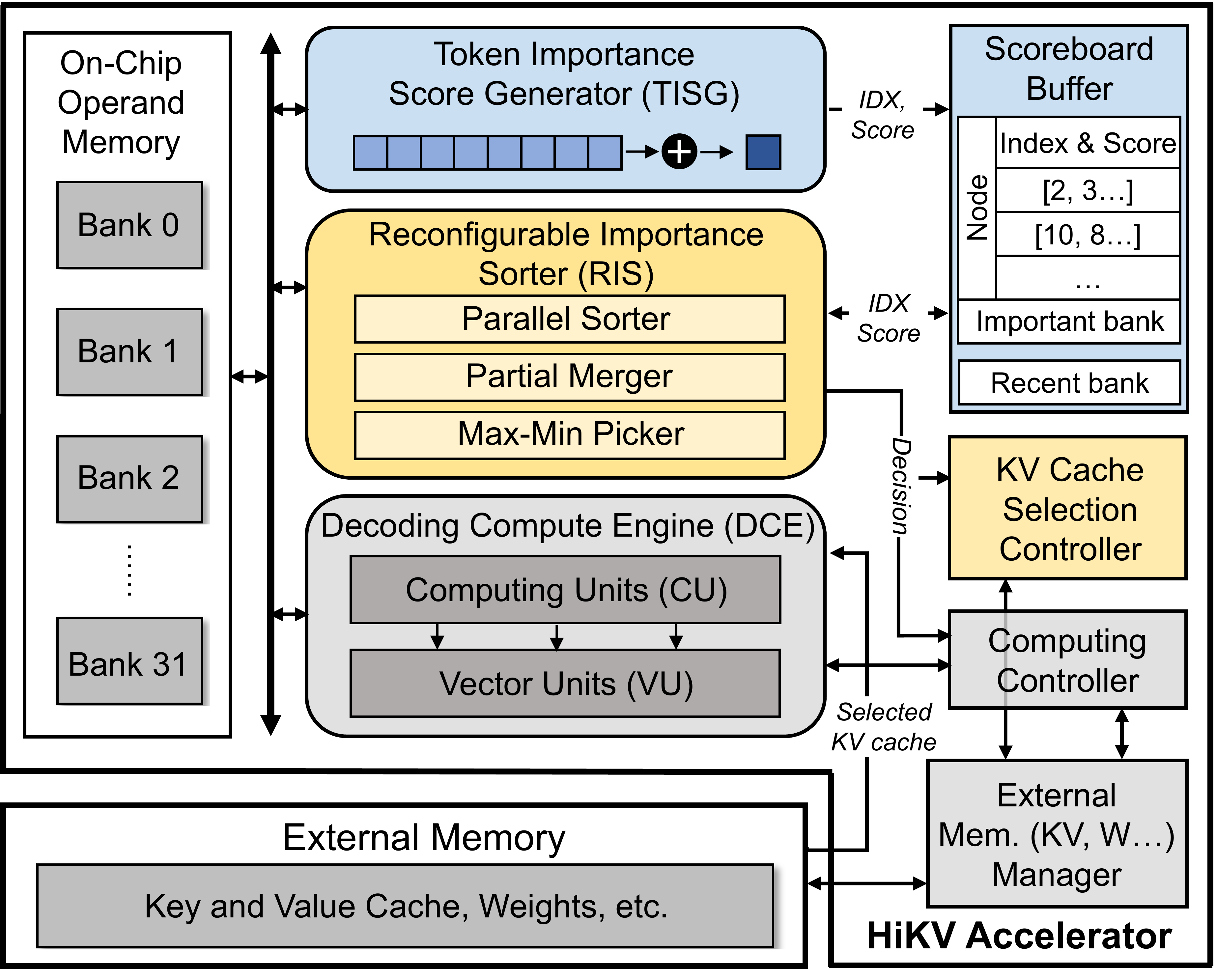}
\caption{\fcRfour{\projName accelerator architecture, where the highlighted components, the RIS, scoreboard buffer, and TISG, constitute the \projName-specific overhead that enables hierarchical two-stage KV cache optimization.}}
\label{fig:archi}
\vspace{-0.4cm}
\end{figure} 

\subsection{System Architecture Overview}

\fcRfour{The overall architecture of the HiKV accelerator is depicted in Fig.~\ref{fig:archi}, which comprises five major functional blocks: the token importance score generator (TISG), the reconfigurable importance sorter (RIS), the scoreboard buffer, the decoding compute engine (DCE), and the KV cache selection controller. These modules are coordinated by a central computing controller and an external memory manager to maintain dataflow during the autoregressive decoding phase.}
\fcRfive{Specifically, TISG is exclusively activated during Stage I importance maintenance. Upon completion of $QK^T$ computation, it reads $P_t$ directly from the on-chip operand memory and updates the IS of each token in the recent bank via 512 parallel FP16 adders. The resulting ISs are retained in Scoreboard. The DCE comprises 512 FP16 MAC units in CU and 64 FP32 floating-point units in VU, responsible for GeMV and non-linear operations, respectively.}
\fcRfour{The RIS serves as the core innovation of \projName design and is elaborated in the following subsections.}

\fcRfour{The on-chip operand memory is organized as 32 independent banks, providing unified on-chip storage for intermediate data throughout the decoding process, including loaded KV cache tokens, query vectors, attention score vectors.}
\fcRfour{The scoreboard buffer maintains two logical regions. The recent token bank is implemented as a FIFO structure that tracks the index and accumulated IS of the most recent $B_I/2$ tokens.}
\fcRfour{The important token bank is a min-heap-based register array that retains the index and frozen IS of the $B_I/2$ tokens with historically significant ISs.}
\fcRfour{By preserving only token indices and their corresponding accumulated or frozen IS, the scoreboard acts as a lightweight metadata structure to guide selective cache access, while the full KV tensors reside exclusively in external memory.}


\fcRfive{During attention computation, the RIS is invoked three times to drive the two-stage importance selection, operating sequentially in Stage~II, Stage~I, and Stage~II modes within a single decoding step, as described in Algorithm~\ref{algo:hikv}.}
\fcRfive{For the $QK^T$ computation, the RIS first performs top-$B_{II}$ sorting on $Q_t$ in Stage~II mode, and the resulting index mask is forwarded to the KV cache selection controller.}
\fcRfive{Concurrently, the token indices maintained in the scoreboard guide the external memory manager to selectively load the corresponding K cache entries.}
\fcRfive{After the VU produces $P_t$ via softmax, the TISG updates the IS registers accordingly, and when the recent bank FIFO ejects its oldest token, the RIS switches to Stage~I mode to perform heap maintenance.}
\fcRfive{For the $PV$ computation, the RIS switches back to Stage~II mode, performing parallel top-$B_{II}$ sorting across all chunks of $P_t$, whose output masks guide the selective loading of V cache entries.}
\fcRfive{For computations outside the attention mechanism, the RIS and TISG are bypassed.}

\begin{figure}[t]
\centering
\includegraphics[width=\columnwidth]{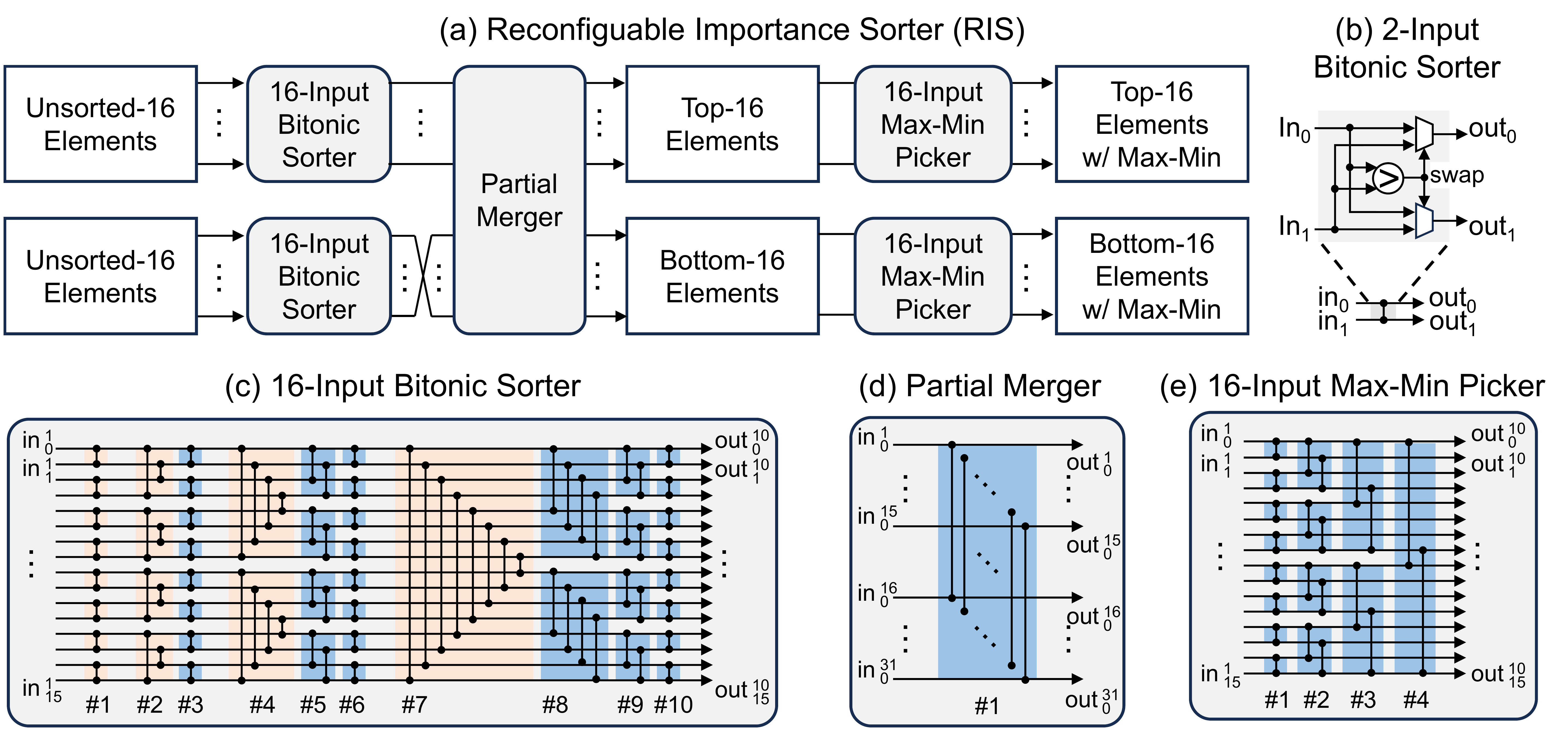}
\caption{\fcRfour{(a) RIS architecture comprises three reconfigurably combined modules, all built upon (b) a unified 2-input bitonic sorter: the parallel sorter centered on (c) a 16-input bitonic sorter, (d) partial merger, and (e) max-min picker.}}
\label{fig:ris}
\end{figure} 

\subsection{\fcRfour{Reconfigurable Importance Sorter (RIS)}}
\fcRfour{The RIS is the core innovation of the HiKV accelerator, comprising three core functional modules as illustrated in Fig.~\ref{fig:ris}(a): the parallel sorter, the partial merger, and the max-min picker.}
\fcRfour{These modules share the 2-input bitonic sorter as a common basic building block, as shown in Fig.~\ref{fig:ris}(b), which contains a comparator and a swap unit that routes the larger and smaller inputs to designated outputs based on the comparison result.}
\fcRfour{Built upon this basic unit, the aforementioned modules are combined in different topological configurations to form the complete RIS circuit, with a critical path of 15 parallel comparison stages when fully activated.}
\fcRfour{Owing to the unified 2-input bitonic sorter building block, the input scale of the RIS can be flexibly extended to accommodate different application scenarios.}

\fcRfive{A 16-input bitonic sorter, as shown in Fig.~\ref{fig:ris}(c), sorts 16 unordered inputs into an ordered sequence and is organized as 80 dual-input bitonic sorters following the topology of a bitonic sorting network~\cite{batcher1968sorting}, corresponding to 10 parallel comparison stages each containing 8 parallel comparators.}
\fcRfour{The parallel sorter comprises two such 16-input bitonic sorters operating in parallel, enabling simultaneous processing of two independent 16-element input vectors.}
\fcRfour{A direction switcher is attached to the output of the parallel sorter.}
\fcRfour{It selects, via a control signal, whether to reverse the output ordering, converting an ascending sequence into a descending one.}

\fcRfour{The partial merger, as shown in Fig.~\ref{fig:ris}(d), accepts two 16-element sorted sequences as input: one in ascending order and one in descending order.}
\fcRfour{Through a single parallel comparison stage, it performs element-wise comparison and redistribution between the two sequences, routing the global top-16 elements to the upper output and the bottom-16 elements to the lower output.}
\fcRfour{Notably, elements within each output group remain unordered, a deliberate design choice that avoids the additional overhead of complete intra-group sorting.}

\fcRfour{The max-min picker processes each group of 16 unordered elements produced by the partial merger, as shown in Fig.~\ref{fig:ris}(e).}
\fcRfour{Built upon the 2-input bitonic sorter, it identifies the minimum and maximum values within each group through 4 parallel comparison stages, placing them at the first and last positions respectively, while remaining elements stay unordered.}
\fcRfour{Retaining intra-group disorder is deliberate, as subsequent heap and selection operations require only the boundary values to be precisely located, rendering complete intra-group ordering unnecessary and thus eliminating the associated circuit overhead.}

\fcRfour{Through differential control signal configurations across the modules, the RIS efficiently supports four operational modes: \texttt{HEAPIFY\_UP} for heap construction during the prefill phase, \texttt{HEAPIFY\_DOWN} for heap maintenance, \texttt{GLOBAL\_SORT} for global sorting of the Q vector, and \texttt{CHUNK\_SORT} for chunk-based sorting of the P vector. Among these, \texttt{HEAPIFY\_UP} and \texttt{HEAPIFY\_DOWN} serve the token-level importance maintenance of Stage~I, while \texttt{GLOBAL\_SORT} and \texttt{CHUNK\_SORT} serve the element-level selection of Stage~II, as elaborated in Sec.~\ref{subsec:stage1} and Sec.~\ref{subsec:stage2}, respectively.}

\begin{figure}[t]
\centering
\includegraphics[width=\columnwidth]{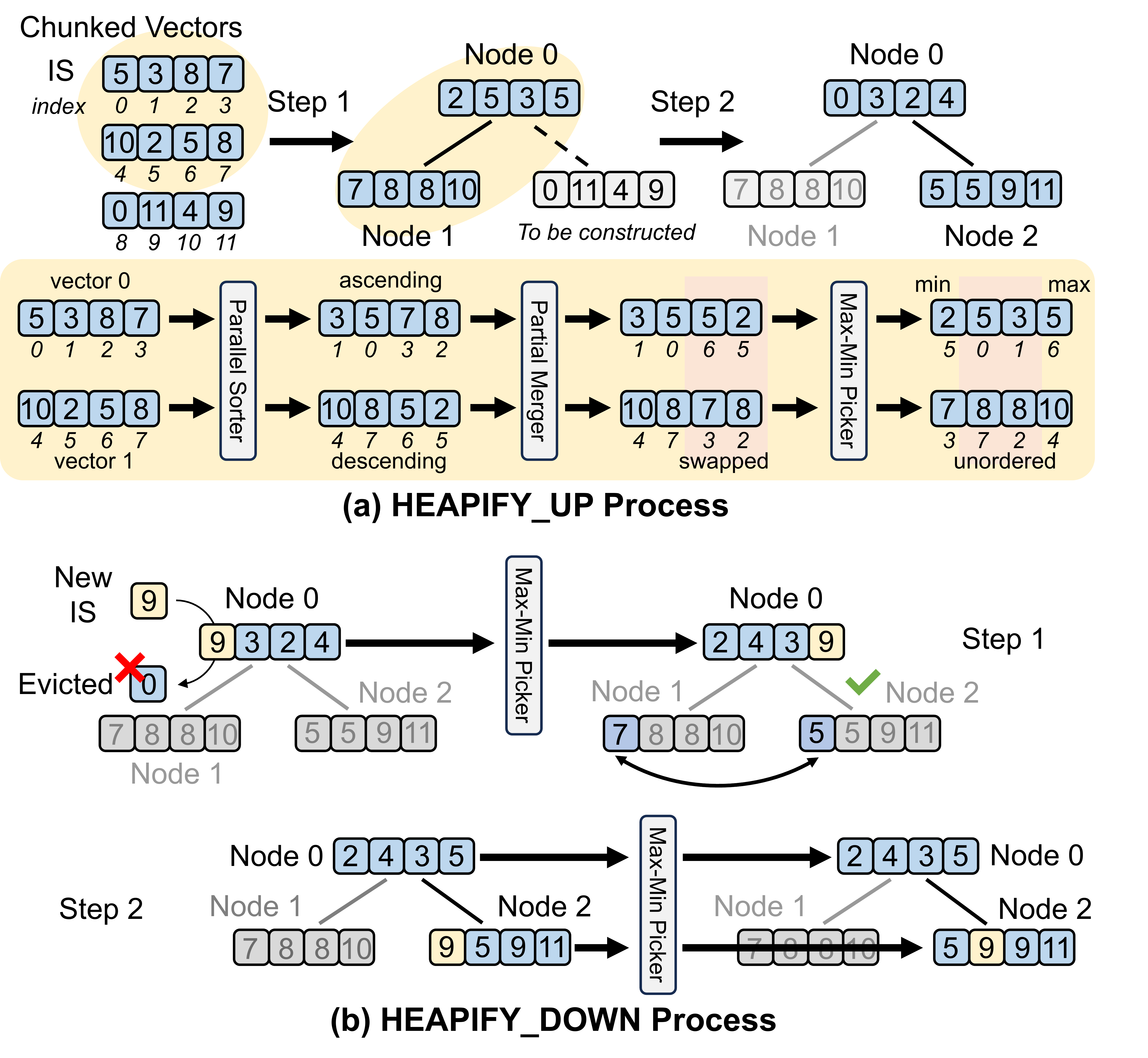}
\caption{\fcRfour{Stage I token importance tracking via RIS-enabled heap operations: (a) \texttt{HEAPIFY\_UP} constructs the vectorized min-heap node by node; (b) \texttt{HEAPIFY\_DOWN} maintains the heap upon token eviction, traversing only a single root-to-leaf path rather than scanning all tokens.}}
\label{fig:stage1}
\end{figure} 

\subsection{Stage I: Heap-Based Token Importance Tracking} \label{subsec:stage1}

\fcRfour{Stage~I invokes two operational modes of the RIS to dynamically maintain the important token bank: \texttt{HEAPIFY\_UP} for initial heap construction at the end of the prefill phase, and \texttt{HEAPIFY\_DOWN} for ongoing heap maintenance.}
\fcRevROne{These operations, together with the per-step IS accumulation, are lightweight. Each decoding step adds only a single vector addition for the IS accumulation and a one-path \texttt{HEAPIFY\_DOWN} update, while \texttt{HEAPIFY\_UP} is a one-time prefill cost amortized over decoding.}

\fcRfour{The RIS employs a parallel comparison network that enables multiple element pairs within a single parent-child node comparison to be processed in parallel, thereby increasing the parallelism and execution efficiency of heap operations.}
\fcRfour{To illustrate the detailed execution flow of both modes, the following examples adopt a 4-way parallel configuration for clarity. Each node in the heap stores a vector of 4 IS along with their corresponding token indices, denoted as $[\cdot]_{(\cdot)}$ where the subscript represents
the token indices.}
\fcRfour{For node $i$ in the important bank, its left and right children are indexed as $2i+1$ and $2i+2$, respectively, and its parent is indexed as $\lfloor(i-1)/2\rfloor$.}
\fcRfour{Note that this vectorized heap requires that the IS$_{\text{max}}$ of any parent node does not exceed the IS$_{\text{min}}$ of its children, with IS$_{\text{min}}$ and IS$_{\text{max}}$ placed at the first and last positions of each node vector, respectively, while the remaining elements stay
unordered.}

\fcRfour{During the \texttt{HEAPIFY\_UP} phase, the RIS fully activates all modules to construct the parent-child hierarchy of the heap in a node-by-node manner.}
\fcRfour{As illustrated in Fig.~\ref{fig:stage1}(a), taking the initial comparison of two candidate nodes $[5, 3, 8, 7]_{(0, 1, 2, 3)}$ and $[10, 2, 5, 8]_{(4, 5, 6, 7)}$ as an example: both nodes are fed into the parallel sorter, where the first is configured via the direction switcher to produce an ascending output $[3, 5, 7, 8]_{(1, 0, 3, 2)}$, and the second a descending output $[10, 8, 5, 2]_{(4, 7, 6, 5)}$.}
\fcRfour{The two sorted outputs are then forwarded to the partial merger, which performs element-wise comparison and redistribution in a single parallel comparison stage, yielding a smaller-element group $[3, 5, 5, 2]_{(1, 0, 6, 5)}$ and a larger-element group $[10, 8, 7, 8]_{(4, 7, 3, 2)}$.}
\fcRfour{Both outputs are subsequently processed by the max-min picker, which identifies the minimum and maximum values within each group, producing Node~0 $[2, 5, 3, 5]_{(5, 0, 1, 6)}$ with IS$_{\text{min}}{=}2$ and IS$_{\text{max}}{=}5$, and Node~1 $[7, 8, 8, 10]_{(3, 7, 2, 4)}$ with IS$_{\text{min}}{=}7$ and IS$_{\text{max}}{=}10$.}
\fcRfour{As each new node is inserted into the heap, the above comparison process is repeated bottom-up against its parent node, until the IS$_{\text{min}}$ of the current node is no less than the IS$_{\text{max}}$ of its parent, completing the hierarchical heap construction.}

\fcRfour{During the \texttt{HEAPIFY\_DOWN} phase, whenever the recent bank FIFO ejects its oldest token, it must be determined whether IS$_{\text{new}}$ of that token should replace an existing element in the important bank heap.}
\fcRfour{Since the positions of IS$_{\text{min}}$ and IS$_{\text{max}}$ within the root node are already known, a state machine rapidly branches into one of two cases.}
\fcRfour{When IS$_{\text{new}}$ $<$ IS$_{\text{min}}$, the RIS is not invoked, and the token is discarded directly.}
\fcRfour{When IS$_{\text{new}}$ $>$ IS$_{\text{min}}$, \texttt{HEAPIFY\_DOWN} is triggered, as illustrated in Fig.~\ref{fig:stage1}(b).}
\fcRfour{Taking IS$_{\text{new}}{=}9_{(12)}$ as an example. It is compared against Node~0 $[0, 3, 2, 4]_{(8, 1, 5, 10)}$, satisfying IS$_{\text{new}}$ $>$ IS$_{\text{min}}$.}
\fcRfour{IS$_{\text{new}}{=}9_{(12)}$ first replaces IS$_{\text{min}}$ in Node~0, yielding $[9, 3, 2, 4]_{(12, 1, 5, 10)}$. Only the max-min picker is then activated to relocate the boundary elements, with the parallel sorter and partial merger both bypassed, updating Node~0 to
$[2, 3, 4, 9]_{(5, 1, 10, 12)}$.}
\fcRfour{The heap property is then checked by comparing Node~0's IS$_{\text{max}}$ against the IS$_{\text{min}}$ values of its children: since Node~1 has IS$_{\text{min}}{=}7$ and Node~2 has IS$_{\text{min}}{=}5$, the right child (Node~2) is selected as the propagation direction.}
\fcRfour{As Node~0's IS$_{\text{max}}{=}9$ $>$ Node~2's IS$_{\text{min}}{=}5$, a heap violation is detected.}
\fcRfour{Node~2's IS$_{\text{min}}$ therefore replaces Node~0's IS$_{\text{max}}$, updating Node~0 to $[2, 3, 4, 5]_{(5, 1, 10, 0)}$, while IS$_{\text{new}}{=}9_{(12)}$ is placed at Node~2's former IS$_{\text{min}}$ position, updating Node~2 to $[9, 5, 9, 11]_{(12, 5, 11, 9)}$; the parallel sorter and partial merger are again bypassed, and only the max-min picker is activated to process Node~2.}
\fcRfour{This process iterates down the heap until the IS$_{\text{max}}$ of the current node does not exceed the IS$_{\text{min}}$ of either of its children, at which point heap maintenance is complete.}

\subsection{Stage II: Chunk-Based Element Importance Selection} \label{subsec:stage2}

\fcRfour{Stage~II performs global sorting over the Q vector within the attention head dimension, and chunk-wise sorting over the P vector using the head dimension as the chunk size, extracting the top-$B_{II}$ most important elements from each chunk to drive selective loading from
external memory.}
\fcRfour{To this end, the RIS supports two operational modes: \texttt{GLOBAL\_SORT} for globally sorting the Q vector prior to $QK^T$ computation, and \texttt{CHUNK\_SORT} for independently sorting each chunk of the P vector prior to $PV$ computation.}
\fcRfour{Since both modes identically invoke the RIS, the following description applies to both. In this configuration, the RIS activates the parallel sorter and the partial merger, while the max-min picker is bypassed.}
\fcRfour{For input P vectors that do not fill a complete chunk, the missing elements are padded with negative infinity to ensure they are excluded from the top-$B_{II}$ selection.}

\begin{figure}[t]
\centering
\includegraphics[width=\columnwidth]{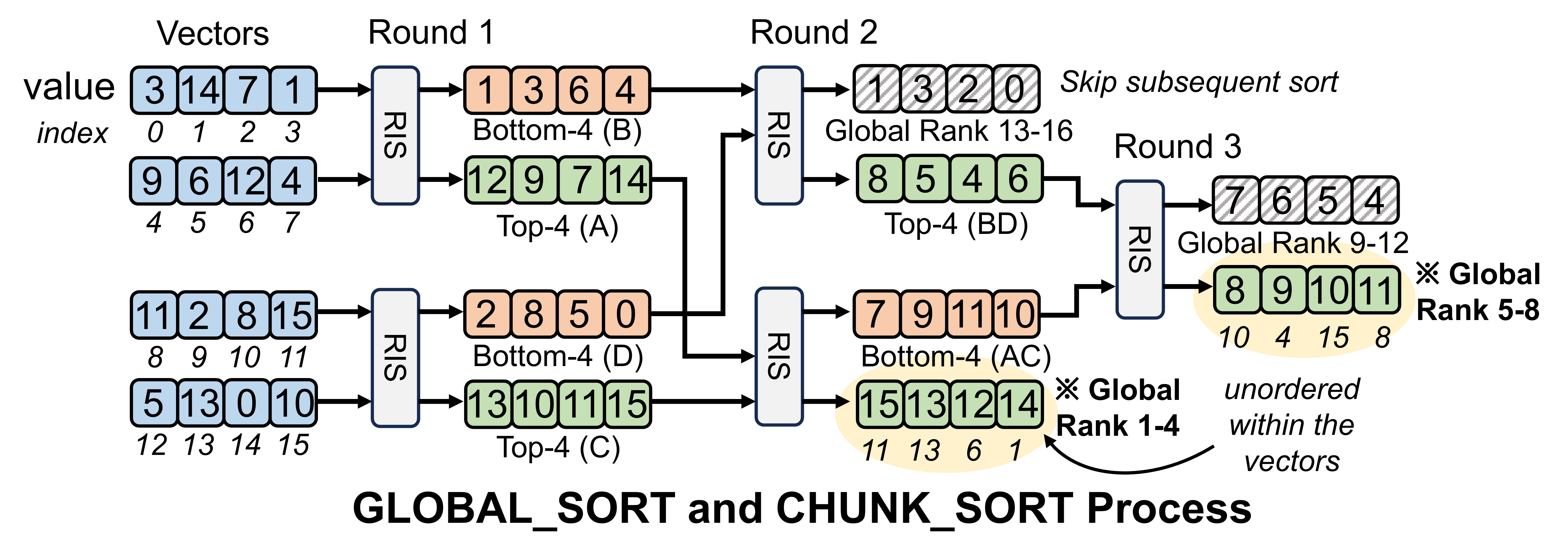}
\caption{\fcRfour{Stage II element importance selection via RIS-enabled chunk-based sorting: \texttt{GLOBAL\_SORT} and \texttt{CHUNK\_SORT} proceed over multiple rounds of RIS invocations, skipping intervals confirmed outside the top-k range early.}}
\label{fig:stage2}
\end{figure} 

\fcRfour{As illustrated in Fig.~\ref{fig:stage2}, and following the same convention as Sec.~\ref{subsec:stage1}, the following example adopts a 4-way parallel configuration for clarity.}
\fcRfour{Taking the extraction of the top-8 elements from 16 inputs as an example, the process completes in three rounds comprising five RIS invocations in total.}
\fcRfour{In Round~1, the 16 elements are partitioned into four groups of 4, which are paired and processed in two parallel RIS invocations, yielding Top-4(A) and Bottom-4(B) from the first pair, and Top-4(C) and Bottom-4(D) from the second.}
\fcRfour{In Round~2, Top-4(A) and Top-4(C) are fed into the RIS, producing the global Top-4 and Mid-4(AC); Bottom-4(B) and Bottom-4(D) are similarly processed, yielding Mid-4(BD) and the global Bottom-4.}
\fcRfour{In Round~3, Mid-4(AC) and Mid-4(BD) are fed into the RIS, producing the elements ranked 5th--8th and those ranked 9th--12th.}
\fcRfour{At this point, all 16 elements are partitioned into four ranked intervals, and the global top-8 is obtained by merging the first two intervals. The corresponding top-$B_{II}$ mask is generated by the KV cache selection controller according to the configured $B_{II}$.}
\fcRfour{In practice, both the input vector dimension and $B_{II}$ are configurable, and the number of RIS invocations adjusts accordingly.}
\fcRevROne{A larger $d_h$ can be handled by increasing the number of invocation rounds, while the RIS needs to be enlarged when the added sorting latency exceeds the external memory access latency and becomes a bottleneck.}

\fcRfour{Stage~II requires only that elements be correctly partitioned into their respective ranked intervals. Once an interval is confirmed to fall entirely outside the top-$B_{II}$ range, it requires no further RIS invocations and is skipped directly.}
\fcRfour{Since boundary element identification within each group is unnecessary, the max-min picker is fully bypassed, reducing unnecessary logic switching and thereby lowering dynamic power consumption.}
\fcRfour{Together with the heap maintenance operations of Stage~I, the same RIS circuit serves the hierarchical KV cache optimization flow through differential control signal configurations, each realizing distinct sorting semantics.}

\subsection{\fcRevROne{Computing Datapath and External Memory Layout}} \label{subsec:datapath}

\begin{figure}[t]
\centering
\fcRevROne{\includegraphics[width=0.9\columnwidth]{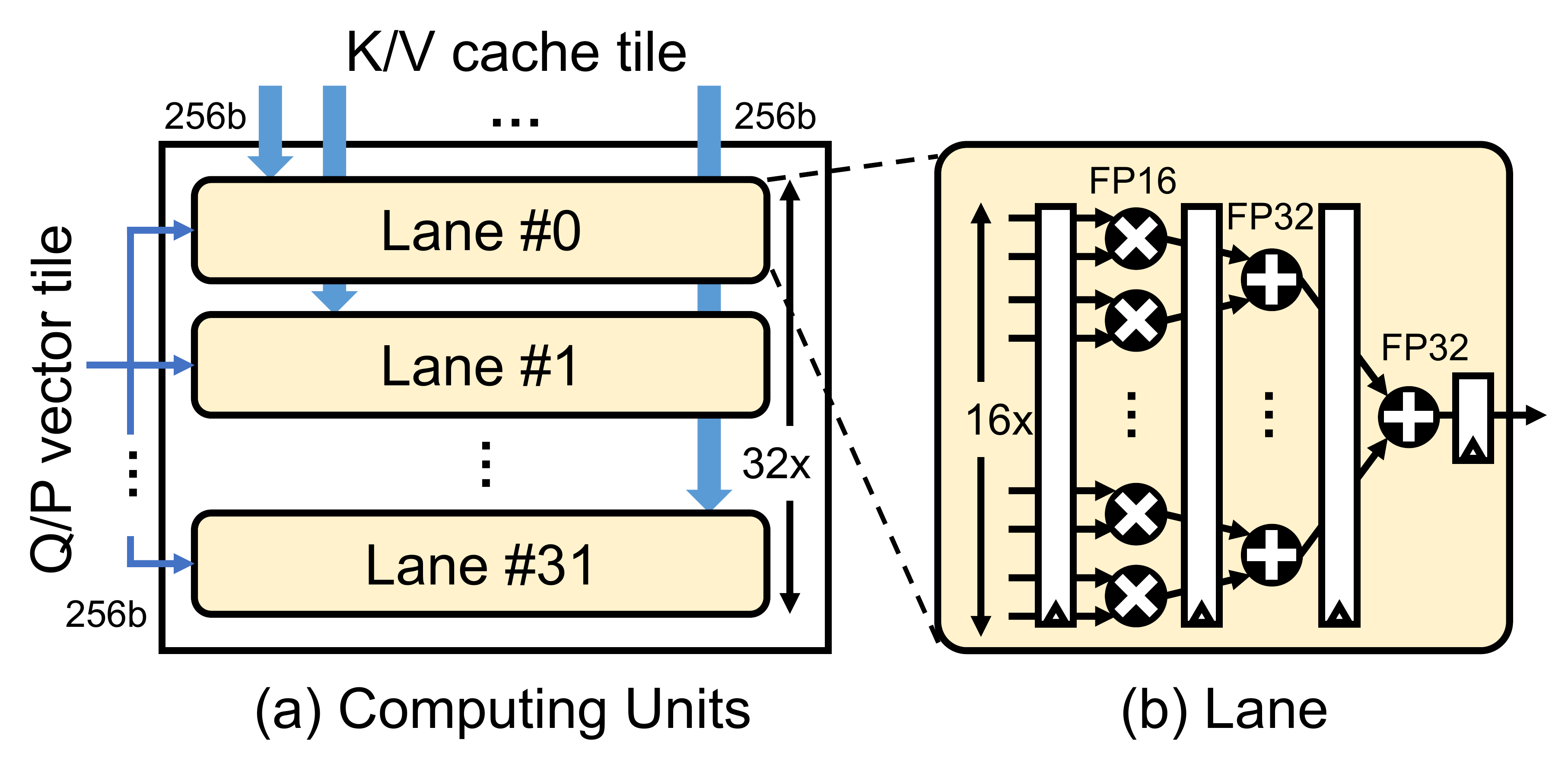}}
\caption{\fcRevROne{The CU comprises (a)~32 parallel lanes, each of which (b)~processes a partial dot-product via 16 FP16 multipliers and an adder tree.}}
\label{fig:dce}
\end{figure}

\fcRevROne{Beyond the RIS, the \projName accelerator also relies on the computing datapath and the external memory layout to complete the execution pipeline.}

\fcRevROne{The sparse index masks produced by the RIS guide the CU in the DCE to execute compressed GeMV over only the selected KV cache entries, reducing both DRAM transfers and multiply-accumulate operations proportionally. As shown in Fig.~\ref{fig:dce}, the CU comprises 32 parallel lanes, each with 16 FP16 multipliers and an adder tree that reduces to a FP32 partial dot-product. One Q or P vector tile of 16 elements is broadcast to all lanes simultaneously, while each lane receives a distinct K or V cache tile of 16 elements as its unicast input. For $QK^T$, the 32 lanes each compute the inner product between $Q_t$ and one K token along the $d_h$ dimension, producing 32 attention logits in parallel. The resulting logits are then passed to the VU, which executes softmax to produce $P_t$. For $PV$, the 32 lanes each accumulate one output element along the $B_I$ token dimension, producing 32 output elements in parallel.}

\begin{figure}[t]
\centering
\fcRevROne{\includegraphics[width=0.92\columnwidth]{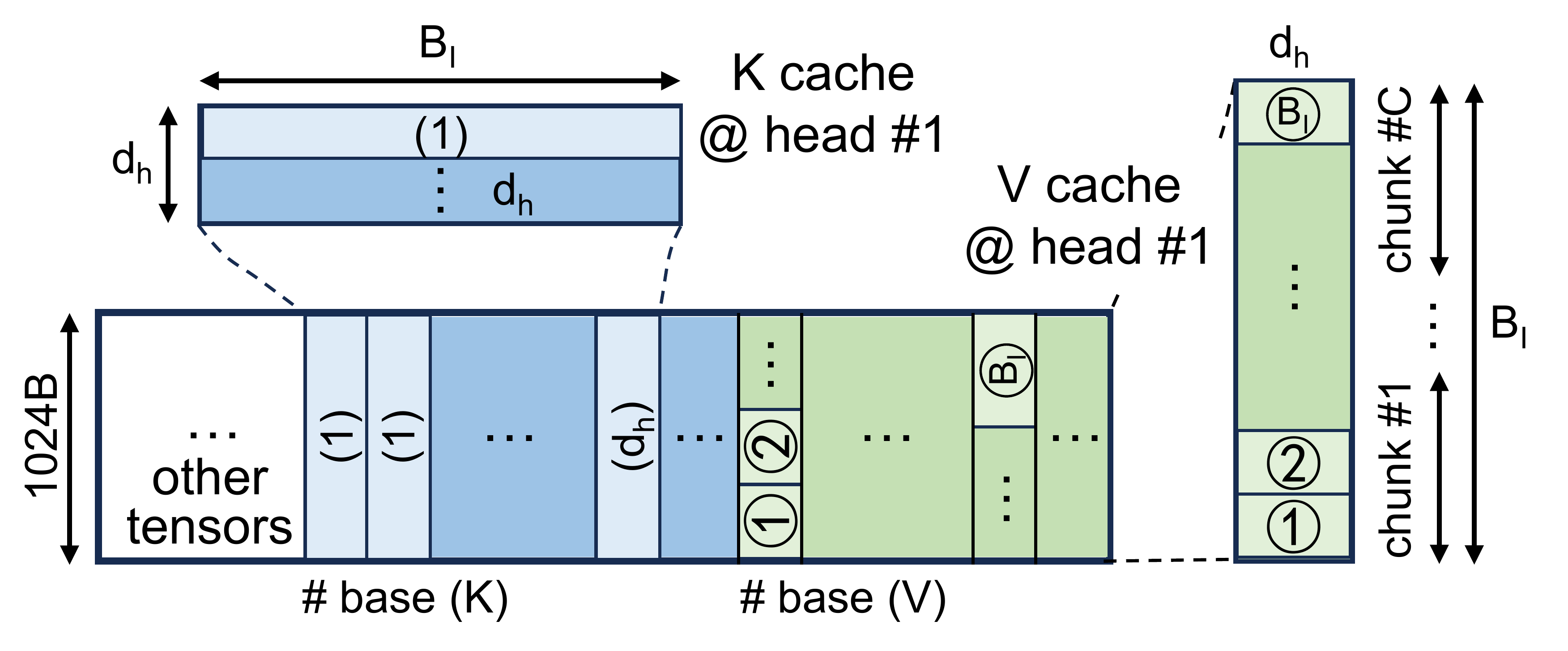}}
\caption{\fcRevROne{Off-chip KV cache layout for one attention head.}}
\label{fig:dram}
\end{figure}

\fcRevROne{To serve the resulting hierarchical access patterns of both stages efficiently, the K and V caches are organized in different DRAM layouts, as shown in Fig.~\ref{fig:dram}. The K cache uses an element-first layout, where all $B_I$ token values for a given element position are stored contiguously. Stage~II's selection of $B_{II}$ element positions therefore translates directly into $B_{II}$ fully contiguous burst transfers. The V cache uses a chunk-indexed layout, where within each of the $C$ chunks, token entries are stored contiguously by token position. Stage~II reads only the $B_{II}$ selected token entries per chunk, and in the worst case these entries span different DRAM rows, causing a row miss that cannot amortize the row-activate latency. Since both stages operate at token-block granularity and Stage~II element-level accesses remain within individual token blocks, the two-stage access pattern is naturally compatible with block-wise memory management schemes such as PagedAttention~\cite{kwon2023efficient}.}

\begin{figure*}[t]
\centering
\includegraphics[width=2\columnwidth]{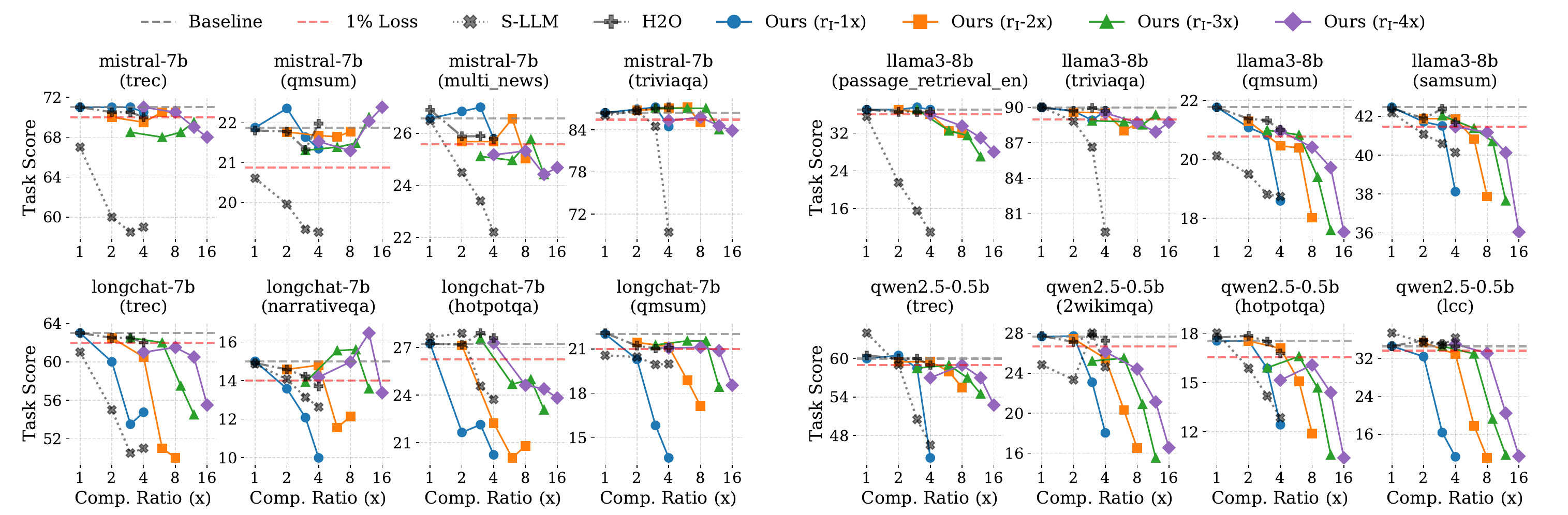}
\caption{Accuracy comparison between \fcRThree{our \projName}, S-LLM~\cite{xiao2024efficient}, and H2O~\cite{zhang2023h2o} under various compression settings for various tasks in LongBench~\cite{bai2024longbench}.}
\label{fig:linechart}
\vspace{-0.3cm}
\end{figure*} 

\section{Experimental Results} \label{sec:eval}

\fcRfour{This section evaluates the \projName co-design across algorithmic, hardware, and system-level dimensions. Following the description of the experimental setup, we first analyze the algorithmic accuracy-compression trade-off. We then quantify the hardware-level area and power efficiency of the implementation. To demonstrate the system-level benefits, we measure the reductions in external memory traffic and improvements in end-to-end inference performance. Finally, we provide a comparative analysis against SotA KV cache accelerators to highlight the superior efficiency of the \projName co-design.}

\subsection{Experimental Setup} \label{subsec:setup}

\textbf{\fcRfour{Algorithmic Evaluation.}}
\fcRfour{We evaluate \projName's algorithmic performance across four representative LLMs: mistral-7b-instruct-v0.2~\cite{jiang2023mistral}, llama3-8b-instruct~\cite{llama3modelcard}, longchat-7b-v1.5-32k~\cite{li2023long}, and qwen2.5-0.5b-instruct~\cite{qwen2}.}
\fcRfour{The evaluation utilizes a selected subset of ten tasks from the LongBench benchmark suite to represent diverse scenarios in question answering, summarization, and reasoning.}
\fcRfour{We fix the prefill prompt length at 4096 tokens and employ single-request inference with a batch size of one to focus on the decoding phase.}
\fcRfour{To comprehensively characterize the accuracy-compression trade-off, we sweep the Stage I token-level compression ratio $r_{I}$ and Stage II element-level compression ratio $r_{II}$ from 1$\times$ to 4$\times$.}
\fcRfour{The total compression ratio is defined as $r_{total} = r_{I} \times r_{II}$.}
\fcRfour{The primary baseline for comparison is the vanilla KV cache implementation, which maintains all KV tensors in FP16 precision without any eviction or selection. We define a performance degradation of less than 1\% compared to this vanilla baseline as the requirement for maintaining algorithmic integrity.}
\fcRfour{We also include SotA importance-based methods such as StreamingLLM (S-LLM)~\cite{xiao2024efficient} and H2O~\cite{zhang2023h2o} to benchmark our hierarchical strategy.}


\textbf{\fcRfour{Hardware Implementation.}}
\fcRfour{The \projName accelerator is implemented in Verilog and synthesized using the Cadence Genus tool. We target the TSMC 16nm technology node with an operating frequency of 300 MHz and a supply voltage of 0.8V. Functional verification is performed using Cadence Xcelium to ensure the correctness of the two-stage sorting and eviction logic across various importance distributions.}
\fcRevROne{The reported area and power are obtained from the post-layout design in Cadence Innovus, with power analysis using activity-annotated post-layout netlists from representative workloads.}
\fcRfour{This allows for a precise breakdown of the power consumed by the \projName-specific components, including the Scoreboard, the RIS, and the TISG. We quantify the area and power overheads of these modules by comparing them against the on-chip computation and memory resources in our \projName accelerator.}

\textbf{\fcRfour{System Simulation.}}
\fcRevROne{We use HBM2 as the external memory interface with a 1024-bit bus width, an average access energy of 3.9 pJ/bit, and a 256 GB/s peak bandwidth following~\cite{chen2025titanus, zhu2025mata, moradifirouzabadi2025end, fang2025anda, jouppi2021ten}.}
\fcRfour{\fcRevROne{The DRAM access is modeled with DRAMSim3~\cite{li2020dramsim3}.} We leverage the LLM-Viewer performance analyzer to model the end-to-end execution of the decoding phase by incorporating the synthesized area and power parameters of the \projName-specific modules. To ensure a fair comparison, we configure a unified hardware platform with 512 FP16 MAC units and 2 MB of on-chip SRAM for all evaluated methods. We implement S-LLM~\cite{xiao2024efficient}, H2O~\cite{zhang2023h2o}, and Token-Picker (Token-P)~\cite{park2024tokenpicker} on this identical platform to eliminate variations arising from disparate hardware assumptions. \fcRfive{For all evaluated methods, each is configured at the most aggressive compression setting that keeps the average score over selected tasks within 1\% of the vanilla baseline. Otherwise, it falls back to the uncompressed baseline.} The systemic energy analysis accounts for both the on-chip computation power and the off-chip memory traffic energy under the previously defined 1\% accuracy constraint. By measuring external memory access reduction, inference speedup, and energy efficiency, we quantify the end-to-end performance gains of our hierarchical pruning strategy across diverse LLM architectures.}

\subsection{Algorithm-Level Evaluation} \label{subsec:acc}

\fcRfour{The algorithmic effectiveness of \projName is validated through an extensive accuracy-compression trade-off analysis across representative LLM architectures.}
\fcRfour{As shown in Fig.~\ref{fig:linechart}, \projName consistently pushes the Pareto frontier of task performance relative to the total KV cache compression ratio $r_{total}$, where $r_{total} = r_{I} \times r_{II}$.}
\fcRfour{In contrast, the SotA importance-aware methods H2O~\cite{zhang2023h2o} and S-LLM~\cite{xiao2024efficient} only facilitate compression at the token-level granularity, where $r_{total}$ is equivalent to $r_I$.}
\fcRfour{This multi-dimensional approach enables \projName to maintain comparable accuracy with the FP16 baseline even under aggressive total compression settings.}
\fcRfour{For instance, with the mistral-7b on trec task, while H2O sustains the 1\% negligible accuracy loss threshold at a $4\times$ compression ratio ($r_{I}=4\times$), \projName leverages its orthogonal Stage II optimization to double the efficiency, achieving an $8\times$ total compression ratio ($r_{I}=4\times, r_{II}=2\times$) under the same accuracy constraint.}
\fcRevROne{This $1\%$ threshold is applied per task, and the attainable ratio thus varies with each task's compression sensitivity, ranging from $2\times$ on sensitive multi-hop reasoning tasks to $16\times$ on compression-tolerant ones.}
\fcRfour{Notably, S-LLM fails to meet this 1\% requirement even at minimal compression levels across several benchmarks.}


\fcRfour{The superior robustness of \projName across diverse tasks stems from the synergistic effect of its two-stage optimization.}
\fcRfour{Unlike the static window-based eviction of S-LLM or the single-granularity dynamic pruning of H2O, HiKV employs a localized accumulation strategy in Stage I to maintain essential tokens, and further eliminates element-level redundancy in Stage II by exploiting the magnitude distribution within vectors.}
\fcRfour{As evidenced by the $r_I$ curves in Fig.~\ref{fig:linechart}, prioritizing token-level significance before element-wise selection yields a more robust performance profile than pushing a single granularity to its extreme. The consistent accuracy gains observed across various LLMs also demonstrate that the hierarchical importance assessment is architecture-agnostic. These results show that \projName preserves the semantic integrity required for complex reasoning while significantly outperforming existing importance-aware compression techniques~\cite{xiao2024efficient, zhang2023h2o}.}
\fcRevROne{We further note that this trade-off can be non-monotonic in a few cases. On LongChat-7B/HotPotQA and Qwen2.5-0.5B/TREC, for example, $(r_I{=}4, r_{II}{=}4)$ exceeds $(r_I{=}1, r_{II}{=}4)$ despite a larger $r_{total}$. This reflects an interaction between the two stages, as tightening Stage~I under aggressive Stage~II concentrates the kept tokens on high-IS entries that still dominate the softmax under the element-level approximation. On most other panels, accuracy decreases monotonically with $r_{total}$.}

\fcRevROne{In our evaluation, $r_I$ and $r_{II}$ are determined by an offline per-task calibration, while both ratios remain runtime-configurable in the \projName accelerator. This renders \projName complementary to dynamic budget allocation methods like Ada-KV~\cite{feng2025adakv} and DynamicKV~\cite{zhou2024dynamickv}, whose adaptively allocated token budget is orthogonal to the two compression dimensions of \projName and can be applied at runtime.}

\begin{table}[t]
\centering
\caption{Component Specifications, Area, and Power Breakdown}
\begin{tabular}{llll}
\toprule
\textbf{Component} & \textbf{Specification} & \textbf{Area (mm\(^2\))} & \textbf{Power (mW)} \\
\midrule
CU           & 512 FP16 MAC          & \fcRevROne{0.141 (6.85\%)}  & \fcRevROne{54.9 (40.5\%)} \\
VU           & 64 × FP32 FPU             & \fcRevROne{0.028 (1.36\%)}  & \fcRevROne{2.01 (1.48\%)} \\
\fcRfive{TISG}         & 512-dim adders            & \fcRevROne{0.009 (0.44\%)}  & \fcRevROne{0.14 (0.10\%)} \\
RIS          & 16-parallel sorter  & \fcRevROne{0.031 (1.51\%)}  & \fcRevROne{1.93 (1.43\%)} \\
Scoreboard    & 8 × 64b × 2K     & \fcRevROne{0.138 (6.70\%)}  & \fcRevROne{7.42 (5.48\%)} \\
Control      & System controls               & \fcRevROne{0.001 (0.05\%)}  & \fcRevROne{0.02 (0.01\%)} \\
Operand Mem.   & 32 × 128b × 4K   & \fcRevROne{1.710 (83.1\%)}  & \fcRevROne{69.0 (51.0\%)} \\
Others       & Others                    & \fcRevROne{0.001 (0.05\%)}  & \fcRevROne{0.01 (0.01\%)} \\
\midrule
\textbf{Total} &                          & \fcRevROne{\textbf{2.059 (100\%)}} & \fcRevROne{\textbf{135.4 (100\%)}} \\
\bottomrule
\end{tabular}
\label{tab:component_breakdown}
\end{table}

\subsection{\fcRfour{Hardware-Level Evaluation}} \label{subsec:hw}

\fcRfour{Table \ref{tab:component_breakdown} summarizes the implementation specifications, area distribution, and power consumption of the \projName architecture.}
\fcRevROne{Implemented in a 16nm CMOS technology node at 300 MHz, the complete \projName accelerator occupies a total silicon area of 2.059 mm$^2$ with a power consumption of 135.4 mW.}
\fcRevROne{The on-chip operand memory dominates the design, taking 83.1\% of the area and 51.0\% of the power, while the computation units occupy 6.85\% of the area but 40.5\% of the power.}

\begin{figure}[t]
\centering
\fcRevROne{\includegraphics[width=0.8\columnwidth]{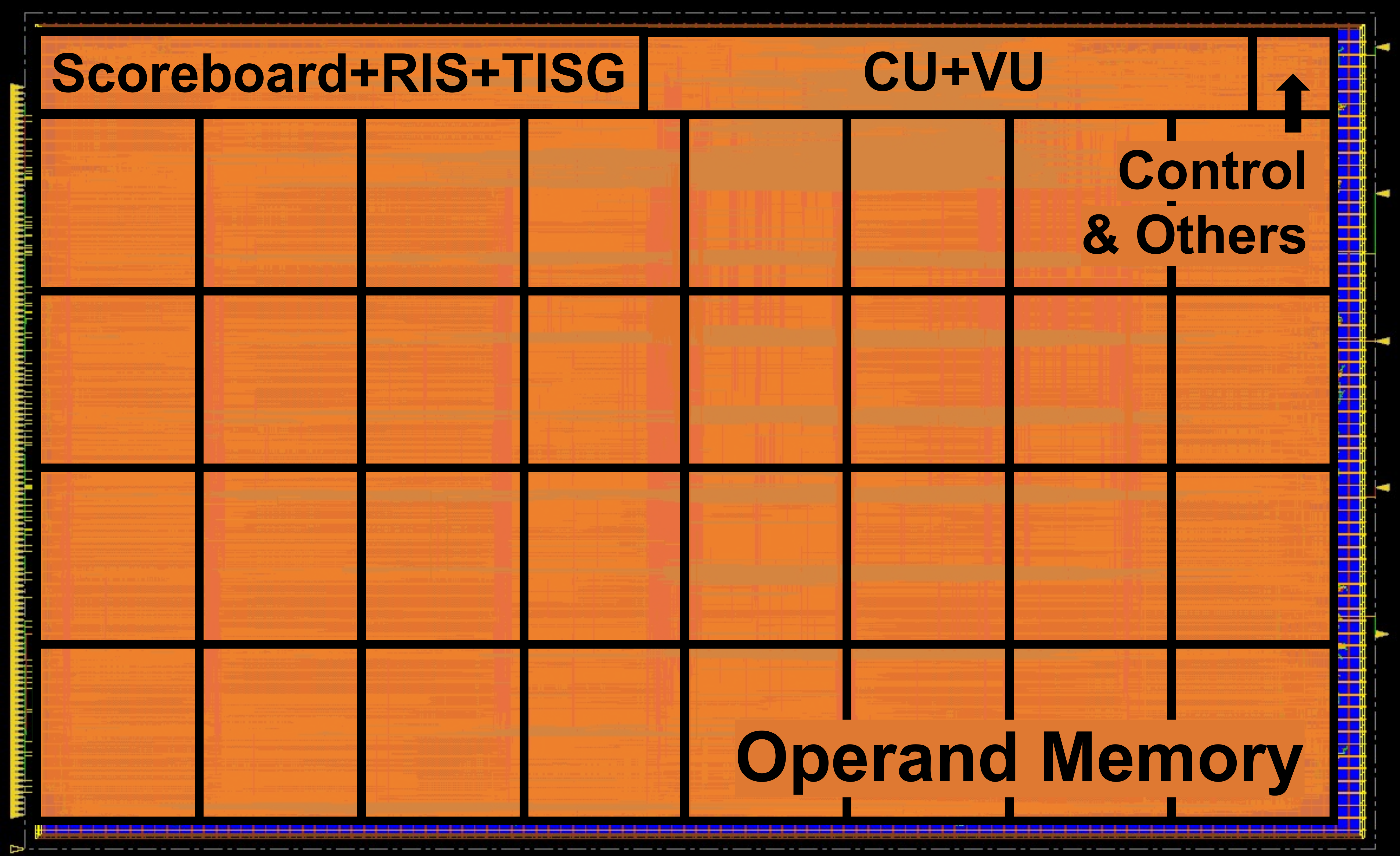}}
\caption{\fcRevROne{Post-layout view of the \projName accelerator in TSMC 16nm.}}
\label{fig:postlayout}
\end{figure}

\fcRfour{Within this resource budget, the hardware overhead specifically introduced by the \projName hierarchical mechanism is minimal.}
\fcRfour{This mechanism consists of the Scoreboard, the RIS, and the TISG, which collectively account for only \fcRevROne{8.64\%} of the total area and \fcRevROne{7.01\%} of the power. This high area efficiency is achieved through the reconfigurable design of the RIS, which utilizes a unified architecture to support four distinct operational modes across both Stage I and Stage II. By enabling hardware-level execution of hierarchical pruning, the specialized \projName architecture aims to minimize the dominant DRAM accesses that characterize conventional LLM decoding.}
\fcRevROne{The corresponding post-layout view is shown in Fig.~\ref{fig:postlayout}.}
\fcRfour{This dedicated hardware implementation provides the basis for the systemic evaluations in the following subsection.}

\subsection{\fcRfour{System-Level Evaluation}} \label{subsec:sys}

\begin{figure}[t]
\centering
\fcRevROne{\includegraphics[width=\columnwidth]{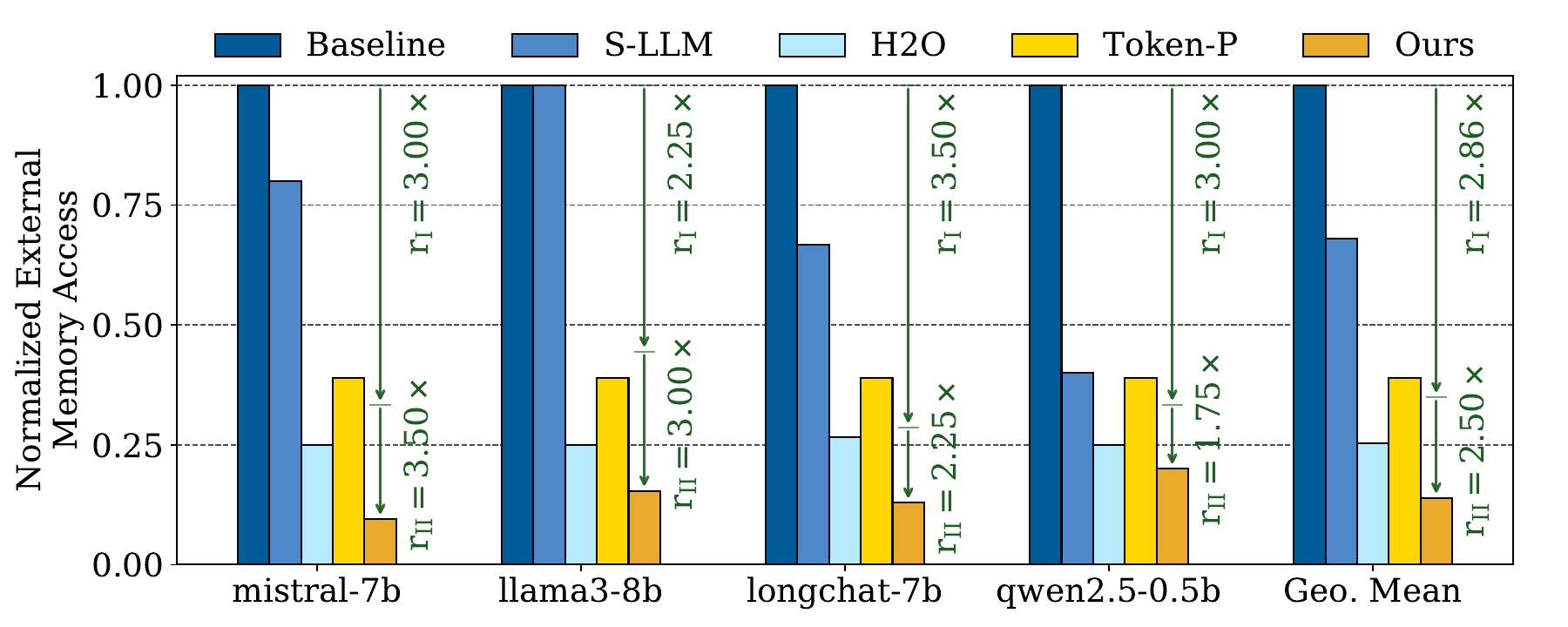}}
\caption{\fcRevROne{External memory access normalized to the vanilla baseline. The \projName reduction is decomposed into the Stage~I ($r_I$) and Stage~II ($r_{II}$) factors.}}
\label{fig:mem}
\end{figure}

\textbf{External Memory Access Reduction.} 
\fcRfour{As shown in Fig.~\ref{fig:mem}, \projName consistently achieves the lowest memory traffic across all tested scenarios under a strict 1\% iso-accuracy constraint. On average, \projName reduces external memory accesses to 13.94\% of the vanilla baseline, significantly outperforming H2O~\cite{zhang2023h2o} (25.41\%), Token-P~\cite{park2024tokenpicker} (38.91\%), and S-LLM~\cite{xiao2024efficient} (67.97\%).}
\fcRfour{For instance, on mistral-7b, \projName compresses traffic to 9.52\%, whereas H2O, Token-P, and S-LLM are limited to 25.0\%, 38.91\%, and 80.0\%, respectively.}
\fcRfour{The inefficiency of S-LLM is attributed to its sliding window policy.}
\fcRfive{On llama3-8b model, it fails to satisfy the accuracy constraint at any compression level, as shown in Fig.~\ref{fig:linechart}, and thus contributes no memory access reduction on that model.}
\fcRfour{In contrast, \projName breaks the single-granularity bottleneck of existing dynamic token-level pruning techniques, like H2O and Token-P, by exploiting element-level redundancy in Stage II. By selectively transferring only significant elements rather than complete vectors, \projName effectively resolves the memory bandwidth bottleneck to enable the substantial speedup observed in the following evaluation.}
\fcRevROne{This reduction can be attributed to the two orthogonal stages, as annotated on the \projName bar in Fig.~\ref{fig:mem}. The token-level Stage~I contributes a reduction of $r_I$ = 2.86$\times$ and the element-level Stage~II contributes $r_{II}$ = 2.50$\times$ on geometric average, which jointly bring the overall reduction to 7.17$\times$. Each stage contributes a substantial portion on its own, and the two prunings are largely complementary.}

\begin{figure}[t]
\centering
\fcRevROne{\includegraphics[width=1\columnwidth]{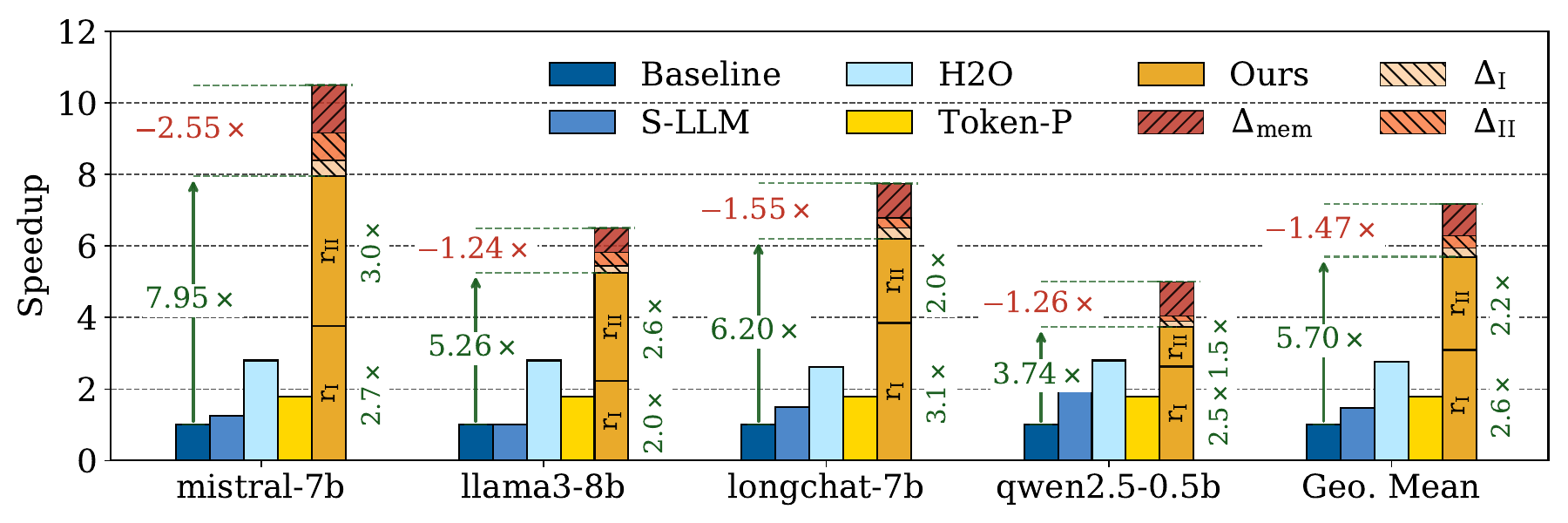}}
\caption{\fcRevROne{Inference speedup normalized to the vanilla baseline. For \projName, the green value is the achieved speedup, split into the Stage~I ($r_I$) and Stage~II ($r_{II}$) factors, and the red value is the overhead, which comes from the memory penalty $\Delta_{\mathrm{mem}}$ and the sorting cost $\Delta_I,\Delta_{II}$.}}
\label{fig:speed}
\end{figure} 

\begin{figure}[t]
\centering
\fcRevROne{\includegraphics[width=1\columnwidth]{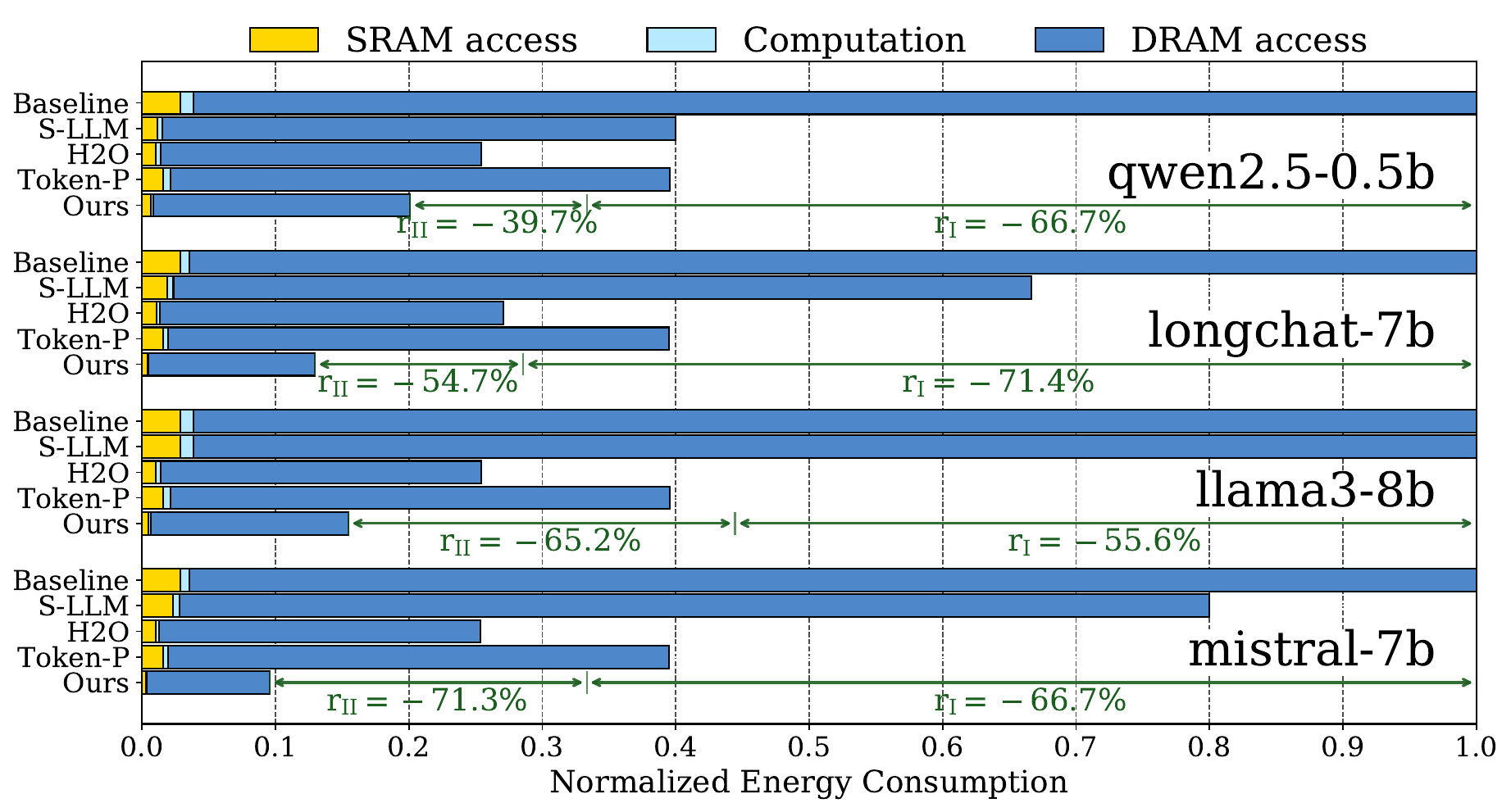}}
\caption{\fcRevROne{Energy consumption normalized to the vanilla baseline, split into DRAM, computation, and SRAM. The \projName reduction is decomposed into the Stage~I ($r_I$) and Stage~II ($r_{II}$) factors.}}
\label{fig:en}
\end{figure} 

\textbf{Inference Speedup.} 
\fcRevROne{As depicted in Fig.~\ref{fig:speed}, \projName delivers an average speedup of approximately 5.70$\times$ across the four representative LLMs, consistently outperforming H2O~\cite{zhang2023h2o} (2.76$\times$), Token-P~\cite{park2024tokenpicker} (1.80$\times$), and S-LLM~\cite{xiao2024efficient} (1.47$\times$). For instance, on mistral-7b, \projName achieves a peak speedup of 7.95$\times$, whereas H2O, Token-P, and S-LLM are limited to 2.80$\times$, 1.80$\times$, and 1.25$\times$, respectively.}
\fcRevROne{The performance gains of H2O are partially constrained by the high computational overhead of maintaining global importance scores through frequent sorting operations. Take qwen2.5-0.5b as an example, H2O reduces external memory accesses by 37.5\% relative to S-LLM, yet its speedup improvement is only 12\%, indicating that the sorting overhead substantially offsets the benefit of its memory access reduction. In contrast, \projName achieves a 1.34$\times$ speedup over H2O, despite only a 1.25$\times$ reduction in memory accesses. This outsized gain highlights the ability of the RIS hardware to more effectively convert memory savings into inference acceleration.}
\fcRfour{By effectively balancing the trade-off between importance maintenance and memory bandwidth, \projName realizes substantial inference speedup compared to the SotAs~\cite{zhang2023h2o, park2024tokenpicker, xiao2024efficient}.}
\fcRevROne{As annotated on the \projName bar in Fig.~\ref{fig:speed}, this speedup also separates into a Stage~I factor of 2.57$\times$ and a Stage~II factor of 2.21$\times$ on geometric average, reaching 5.70$\times$ in total. The speedup lands slightly below the 7.17$\times$ access reduction because of the on-chip sorting and memory access overheads of the accelerator. The dominant overhead is the memory penalty from the fine-grained element-level access of Stage~II, while the RIS sorting cost is only about 8\% and is largely masked by this fine-grained memory access overhead on the critical path.}

\textbf{Energy Consumption.} 
\fcRfour{The systemic energy consumption is analyzed by breaking down the operational costs into DRAM access, SRAM access, and 
computation components.}
\fcRfour{As shown in Fig.~\ref{fig:en}, \projName achieves an energy reduction of 80\% to 90\% relative to the vanilla baseline across all evaluated models. This superior efficiency is primarily driven by the drastic mitigation of off-chip DRAM movement, which accounts for over 96\% of the energy in the vanilla baseline.}
\fcRfour{For instance, on the longchat-7b model, \projName eliminates approximately 87\% of the total energy consumption, whereas H2O~\cite{zhang2023h2o} and Token-P~\cite{park2024tokenpicker} reduce energy by approximately 73\% and 60\%, respectively. These gains are mainly enabled by Stage II element-level selection. On the mistral-7b model, \projName reduces the DRAM energy component by 75.5\% compared to Token-P. This systemic leap is achieved by performing specialized on-chip importance sorting to prune redundant data before transmission, investing a minimal \fcRevROne{7.01\%} hardware power overhead. By leveraging efficient on-chip processing to minimize energy-expensive external memory transfers, \projName provides a promising KV cache co-design method for energy-efficient LLM decoding.}
\fcRevROne{Looking at the two stages, this energy saving is delivered by the two prunings jointly cutting the dominant DRAM energy. On mistral-7b, Stage~I alone reduces the DRAM energy by 66.7\% and Stage~II alone by 71.4\%, and as the two act multiplicatively rather than additively, together they cut 90.5\%. Since the on-chip computation and SRAM that include the RIS account for only about 4\% of the total energy, this algorithmic external memory access saving is carried over almost directly into the end-to-end energy reduction.}
\begin{table*}[!t]
\centering
\caption{Comparison with State-of-the-Art KV Cache Accelerators}
\label{tab:sota}
\renewcommand{\arraystretch}{1.2}
\resizebox{\textwidth}{!}{%
\begin{threeparttable}
\begin{tabular}{lcccccccc}
\toprule
& \textbf{Token-Picker~\cite{park2024tokenpicker}}
& \textbf{Titanus~\cite{chen2025titanus}}
& \textbf{UniCAIM~\cite{xu2025unicaim}}
& \textbf{KV-MMU~\cite{moradifirouzabadi2025end}}
& \textbf{MATA~\cite{zhu2025mata}}
& \textbf{VEDA~\cite{wang2025veda}}
& \textbf{HiKV (Ours)} \\
\midrule
\textbf{Venue / Year}
& DAC'24
& GLSVLSI'25
& DAC'25
& JETCAS'25
& TC'25
& DAC'25
& TCAS-I'26 \\
\textbf{Optimization Granularity}
& Token-level
& Element-level
& Token-level
& Token-level
& Token-level
& Token-level
& \textbf{Hierarchy-level} \\
\textbf{Technology Node (nm)}
& 65
& 14
& 45
& 65
& 28
& 28
& \textbf{16} \\
\textbf{Clock Frequency (MHz)}
& 500
& 200
& N/A
& 800
& 1000
& 1000
& \textbf{300} \\
\fcRevROne{\textbf{Implementation}}
& \fcRevROne{Pre-layout}
& \fcRevROne{Pre-layout}
& \fcRevROne{N/A}
& \fcRevROne{Pre-layout}
& \fcRevROne{Pre-layout}
& \fcRevROne{Pre-layout}
& \fcRevROne{\textbf{Post-layout}} \\
\textbf{Area Consumption (mm\textsuperscript{2})}\fcRevROne{\tnote{~(5)}}
& 8.59\fcRevROne{$^{[0.45]}$}
& 83.32\fcRevROne{$^{[112.3]}$}
& N/A
& $\sim$49.80\fcRevROne{$^{[2.63]}$}\tnote{~(1)}
& 4.601\fcRevROne{$^{[1.42]}$}
& 1.058\fcRevROne{$^{[0.33]}$}
& \fcRevROne{\textbf{2.059}} \\
\textbf{Power Consumption (mW)}\fcRevROne{\tnote{~(5)}}
& 3149\fcRevROne{$^{[1136]}$}
& 31{,}950\fcRevROne{$^{[36{,}016]}$}
& N/A
& N/A
& 461.4\fcRevROne{$^{[297.2]}$}
& 375.3\fcRevROne{$^{[241.8]}$}
& \fcRevROne{\textbf{135.4}} \\
\textbf{Arithmetic Precision}
& INT12
& INT2--3 (mixed)
& Analog+10b ADC
& INT8
& FP16
& FP16
& \textbf{FP16} \\
\textbf{Energy Efficiency (TOPS/W)\tnote{\fcRevROne{~(2),(5)}}}
& 2.78\fcRevROne{$^{[7.70]}$}
& N/A
& N/A
& 2.59\fcRevROne{$^{[7.18]}$}
& N/A
& 0.65\fcRevROne{$^{[1.01]}$}
& \fcRevROne{\textbf{2.26}} \\
\textbf{External Memory Access Reduction\tnote{~(3)}}
& 2.57$\times$
& 2.4$\times$
& N/A
& 2.6$\times$
& $\sim$4.09$\times$\tnote{~(4)}
& 5$\times$
& \textbf{7.17$\times$} \\
\bottomrule
\end{tabular}
\begin{tablenotes}[flushleft]
\footnotesize
\item[(1)] Estimated by summing module-level area breakdowns reported in the original paper.
\item[(2)] Computed based on on-chip synthesized power, excluding off-chip DRAM access energy.
\item[(3)] Reduction factor of off-chip memory access volume relative to the vanilla full-KV cache baseline.
\item[(4)] Estimated as $1.9\times$ (SpAtten vs.\ full-KV baseline) $\times$ $2.15\times$ (MATA vs.\ SpAtten, geomean) $\approx 4.09\times$.
\fcRevROne{\item[(5)] Each entry is shown as ``\emph{reported}$^{[*]}$'', where $^{[*]}$ denotes the value scaled to the 16\,nm node using DeepScaleTool~\cite{sarangi2021deepscaletool}.}
\end{tablenotes}
\end{threeparttable}
}
\end{table*}

\subsection{\fcRfour{Comparison with State-of-the-Arts}} \label{subsec:sota}

\fcRfour{Table~\ref{tab:sota} provides a systematic comparison of HiKV against SotA KV cache accelerators across algorithmic granularity and hardware implementation.}
\fcRevROne{For a fair cross-node comparison, the reported area, power, and energy efficiency of all prior arts are normalized to the 16\,nm node of \projName using DeepScaleTool~\cite{sarangi2021deepscaletool}, shown as the $^{[\cdot]}$ superscripts.}
\fcRfour{Since external memory access in the LLM decoding is dominated by KV cache traffic, its reduction serves as the primary efficiency metric for KV cache accelerators, and all comparisons reported below are obtained under 1\% accuracy loss constraint.}
\fcRfour{Token-level methods, including Token-Picker~\cite{park2024tokenpicker}, VEDA~\cite{wang2025veda}, KV-MMU~\cite{moradifirouzabadi2025end}, and MATA~\cite{zhu2025mata}, achieve reductions in the range of 2.57$\sim$5$\times$, reflecting an inherent ceiling of token-granularity compression.}
\fcRfour{These approaches reduce memory traffic by evicting low-importance tokens, yet each retained token needs to be fully loaded with all its elements, leaving intra-token element-level redundancy unaddressed.}
\fcRfour{Although Titanus~\cite{chen2025titanus} incorporates element-level operations, the absence of coarse-grained token-level filtering limits its external memory access reduction to only 2.4$\times$, which falls below several token-only designs, showing that single-granularity compression strategies face fundamental limitations regardless of the targeted granularity.}
\fcRfour{\projName addresses this bottleneck through the reconfigurable design of the RIS, which time-multiplexes two semantically distinct importance sorting requirements within a unified circuit, serving Stage~I token-level ranking and Stage~II element-level ranking in an alternating fashion.}
\fcRfour{This hardware capability enables the compression gains from both stages to accumulate cooperatively, yielding the highest external memory access reduction in Table~\ref{tab:sota}, substantially surpassing VEDA~\cite{wang2025veda} and MATA~\cite{zhu2025mata}.}
\fcRfour{The hardware overhead introduced by this hierarchical compression mechanism, comprising the Scoreboard, the RIS, and the TISG, amounts to only \fcRevROne{8.64\%} of the total area, with the RIS alone occupying merely \fcRevROne{1.51\%}, supporting the full hierarchical algorithm at minimal silicon cost.}
\fcRfour{In summary, \projName is the first design to realize hierarchical granularity compression, and it is precisely the algorithm-hardware co-design methodology that enables this hierarchical compression to be efficiently realized, thereby breaking through the single-granularity bottleneck and achieving the highest external memory access reduction of 7.17$\times$ among all prior arts.}

\section{Conclusion} \label{sec:concls}

\fcRfour{This paper presents \projName, an algorithm-hardware co-design framework for LLM decoding that systematically alleviates the memory bottleneck
imposed by the KV cache.}
\fcRfour{Built on the insight that KV cache redundancy exists at both token and element granularities, \projName employs two orthogonally complementary compression stages, supported by the reconfigurable importance sorter as the core hardware innovation that accommodates the semantically distinct sorting requirements of both stages within a unified circuit, incurring only 8\% area and power overhead in total.}
\fcRfour{Under less than 1\% accuracy loss, \projName achieves superior compression efficiency over token-level SotA methods, and attains up to \fcRevROne{7.95}$\times$ inference speedup and 90\% energy reduction at the system level.}
\fcRfour{Under iso-accuracy constraints, \projName delivers an additional 1.82$\sim$4.87$\times$ external memory access reduction over state-of-the-art methods.}
\fcRfour{As size of context windows continues to grow, \projName offers a co-design methodology for energy-efficient LLM decoding under increasing KV cache memory pressure.}
\section*{Acknowledgment} \label{sec:ack}

\fcRevROne{The authors sincerely thank the anonymous reviewers for their constructive feedback, and Xinfa Zheng, Sander Crols, and Robin Geens for their generous technical support.}
\fcRevRTwo{The authors acknowledge EuroHPC JU for awarding the project ID EHPC-DEV-2026D03-124 access to Leonardo BOOSTER hosted by CINECA, Italy, and NVIDIA Corporation for awarding GPU hardware through its Academic Grant Program.}

\ifCLASSOPTIONcaptionsoff
  \newpage
\fi



\normalem
\bibliographystyle{IEEEtran}
\bibliography{ref}
%




\end{document}